\RequirePackage{xcolor}
\documentclass[journal,twoside,web]{ieeecolor} 

\usepackage{etoolbox}
\makeatletter
\@ifundefined{color@begingroup}%
  {\let\color@begingroup\relax
   \let\color@endgroup\relax}{}%
\def\fix@ieeecolor@hbox#1{%
  \hbox{\color@begingroup#1\color@endgroup}}
\patchcmd\@makecaption{\hbox}{\fix@ieeecolor@hbox}{}{\FAILED}
\patchcmd\@makecaption{\hbox}{\fix@ieeecolor@hbox}{}{\FAILED}

\usepackage{tmi}
\usepackage{cite}
\usepackage{amsmath,amssymb,amsfonts}
\usepackage{algorithmic}
\usepackage{graphicx}
\usepackage{textcomp}
\usepackage[verbose]{placeins}
\usepackage{float}
\usepackage[switch]{lineno}
\def\BibTeX{{\rm B\kern-.05em{\sc i\kern-.025em b}\kern-.08em
    T\kern-.1667em\lower.7ex\hbox{E}\kern-.125emX}}
\usepackage{anyfontsize}
\usepackage{multirow}
\usepackage[resetcount,lined,boxed,commentsnumbered,ruled,vlined,longend]{algorithm2e}

\makeatletter
\newcommand{\removelatexerror}{\let\@latex@error\@gobble}
\makeatother
\usepackage{array}
\usepackage{fixltx2e}
\usepackage{stfloats}

\usepackage{url}
\usepackage{amssymb}

\usepackage{soul}
\usepackage[clean]{revdiff}

\def\L{{\cal L}}

\def\fhat{\hat{\boldsymbol{\mathrm{f}}}}
\def\v{\boldsymbol{v}}

\def\U{\boldsymbol{U}}

\def\Th{\boldsymbol{\Theta}}

\def\n{\boldsymbol{n}}
\def\g{\boldsymbol{g}}
\def\r{\boldsymbol{r}}

\def\Kb{\boldsymbol{K}}

\def\etal{\textit{et al.\ }}

\def \sc {\mathrm{sc}}

\def \L {\mathcal{L}}

\def \path{\mathbb{P}}

\def \f{\boldsymbol{\mathrm{f}}}
\def \fobj{\boldsymbol{f}}

\begin{document}
\title{DEMIST: A deep-learning-based task-specific denoising approach for myocardial perfusion SPECT}
  
\author{Md Ashequr Rahman, \IEEEmembership{Student Member, IEEE}, Zitong Yu, \IEEEmembership{Student Member, IEEE}, Richard Laforest, Craig K. Abbey, Barry A. Siegel, and Abhinav K. Jha*, \IEEEmembership{Senior Member, IEEE}
\thanks{M. A. Rahman and Z. Yu are with the Department of Biomedical Engineering, Washington
University, St. Louis, MO 63130 USA.
}% <-this % stops a space
\thanks{C. K. Abbey is with the Department of Psychological and Brain Sciences, University of California, Santa Barbara, CA 93106 USA.}% <-this % stops a space
\thanks{R. Laforest and B. A. Siegel are with the Mallinckrodt Institute of Radiology, Washington University, St. Louis, MO
63130 USA.}%
\thanks{*A. K. Jha is with the Department of Biomedical Engineering and
Mallinckrodt Institute of Radiology, Washington University, St. Louis, MO
63130 USA (e-mail: a.jha@wustl.edu).}}
\maketitle
\rlegend
%\vspace{50mm}
\begin{abstract}
There is an important need for methods to process myocardial perfusion imaging (MPI) SPECT images acquired at lower radiation dose and/or acquisition time such that the processed images improve observer performance on the clinical task of detecting perfusion defects. To address this need, we build upon concepts from model-observer theory and our understanding of the human visual system to propose a \underline{De}tection task-specific deep-learning-based approach for denoising \underline{M}P\underline{I} \underline{S}PEC\underline{T} images (DEMIST). The approach, while performing denoising, is designed to preserve features that influence observer performance on detection tasks. We objectively evaluated DEMIST on the task of detecting perfusion defects using a retrospective study with anonymized clinical data in patients who underwent MPI studies across two scanners (N = 338). The evaluation was performed at low-dose levels of 6.25\%, 12.5\% and 25\% and using an anthropomorphic channelized Hotelling observer. Performance was quantified using area under the receiver operating characteristics curve (AUC). Images denoised with DEMIST yielded significantly higher AUC compared to corresponding low-dose images and images denoised with a commonly used task-agnostic DL-based denoising method. Similar results were observed with stratified analysis based on patient sex and defect type. Additionally, DEMIST improved visual fidelity of the low-dose images as quantified using root mean squared error and structural similarity index metric. A mathematical analysis revealed that DEMIST preserved features that assist in detection tasks while improving the noise properties, resulting in improved observer performance. The results provide strong evidence for further clinical evaluation of DEMIST to denoise low-count images in MPI SPECT.
\end{abstract}

%====================================================================================
% introduction
%====================================================================================
\section{Introduction}
Single-photon emission computed tomography (SPECT) myocardial perfusion imaging (MPI) has an established and well-validated role in evaluating patients with known or suspected coronary artery disease \cite{zaret2010clinical}. For diagnosis of this disease, the clinical task performed on MPI-SPECT images is the detection of focally reduced tracer uptake (perfusion defects) reflecting reduced blood flow in the myocardial wall. Typically, in clinical MPI-SPECT protocols, patients are administered a radiopharmaceutical tracer, such as Tc-99m sestamibi or Tc-99m tetrofosmin, under stress and rest conditions. For a protocol involving a Tc-99m radiopharmaceutical with rest and stress imaging performed on a single day, the administered activity can be as high as 48 mCi \cite{henzlova2016asnc}. Thus, developing protocols to reduce this administered dose are well poised for a strong clinical impact \cite{beller2010importance,cohen2010radiation}. Additionally, current MPI-SPECT acquisition protocols can take up to around 12-15 minutes, during which time, the patient is required to be stationary. This is a challenge, especially for older patients, which are a large fraction of the patient population \cite{yuan2015coronary}. Thus, methods to reduce acquisition time can make MPI-SPECT more comfortable for patients, less susceptible to patient motion, and can also lead to increased clinical throughput and reduced cost of imaging. However, reducing this dose and/or acquisition time results in a lower number of detected counts in the projection data, which, when reconstructed, yields images with deteriorated image quality in terms of the ability to reliably detect perfusion defects. Thus, there is an important need to develop methods to process low-count MPI-SPECT images for improved performance on detection tasks.

In recent years, deep learning (DL)-based image-denoising methods have shown promise in predicting normal-dose images from low-dose images for MPI SPECT \cite{ramon2018initial,aghakhan2022deep,sun2022pix2pix,sun2022deep}. Typically, these approaches are trained by minimizing a loss function based on image fidelity, such as pixel-wise mean squared error (MSE) between the actual normal-dose image and low-dose image. These methods have usually been evaluated with fidelity-based metrics such as root mean squared error (RMSE) and structural similarity index metric (SSIM), where the results have indicated that the methods provide improved performance compared to low-dose images. However, it is well recognized that for clinical translation, DL-based denoising methods need to be evaluated on performance in clinically relevant tasks \cite{barrett1990objective,jha2021objective,barrett2013foundations,pretorius2023observer}. At an early stage of translation, model observers provide a mechanism to perform such evaluation \cite{krupinski2021important}. However, of the various DL-based denoising methods proposed for MPI SPECT, those that have been evaluated on the clinical task of detecting perfusion defects have not shown improved performance \cite{yu2023ai,pretorius2023observer}. Recent studies in other imaging modalities have also yielded similar findings \cite{li2021assessing,prabhat2021deep}. 
While it is well recognized that any image-processing method cannot improve the performance of ideal observers due to data-processing inequality \cite{barrett2013foundations}, for sub-optimal observers such as human observers, improving detection performance may be possible. 
Further, the detection task on MPI-SPECT images is clinically performed by human observers. 
Thus, in this manuscript, we investigate the development of denoising methods that explicitly demonstrate improved performance on the task of detecting perfusion defects in MPI-SPECT images with \rchange{human observers}{an anthropomorphic model observer that has been shown to emulate human observer performance on this task \cite{sankaran2002optimum,wollenweber1999comparison}}.

To investigate the limited performance of DL-based denoising methods on detection tasks in MPI SPECT, Yu \etal \cite{yu2023ai} conducted a mathematical analysis with a commonly used DL-based denoising method that used pixel-wise MSE as the loss function. They analyzed the detection performance of a numerical observer that has been observed to emulate human-observer performance in MPI SPECT. Their analysis revealed that the method was improving the noise characteristics of the images, which, in isolation, would have improved observer performance. However, the analysis also indicated that the method was also discarding features used to perform the detection task, which eventually translated to no improvement in observer performance. These observations indicate that a denoising method that can preserve task-specific features may improve observer performance on detection tasks. Recently, in the context of X-ray CT, a few DL-based denoising methods have been proposed with the aim of preserving features that assist in the detection task \cite{ongie2022optimizing,li2022task,han2021low}. These methods typically incorporate a hybrid loss consisting of image fidelity and task-specific terms, where the latter term has been incorporated in the form of signal-to-noise ratio \cite{ongie2022optimizing}, binary cross-entropy loss associated with a DL-based observer \cite{li2022task}, and perceptual loss obtained from features extracted by a pre-trained Visual Geometry Group network \cite{han2021low}. Results from these studies support the idea that preserving task-specific features may assist with improving performance on detection task. However, the methods proposed have limitations to the applicability to the SPECT denoising problem, such as assuming 2D images, defect-known-exactly setups, use of ground-truth phantom as the target/label, and limited interpretability of the task-specific loss term. Additionally, the methods have been evaluated using simulated or stylized studies. For clinical applicability, evaluation of such methods with clinical data and on clinically relevant tasks is needed. 

Motivated by these observations from prior studies, we propose a DL-based task-specific denoising method for 3D MPI SPECT.  The method builds upon concepts from the literature on model observers and our understanding of the human visual system to preserve detection-task-specific features while performing denoising. We objectively evaluate the proposed method on the task of detecting perfusion defects using a retrospective study with anonymized clinical MPI-SPECT data. Additionally, we evaluate the effect of population characteristics, including patient sex and perfusion defect types, on the detection-task performance. Preliminary results of this work have been presented previously \cite{rahman2023task}.

%====================================================================================
% Method
%====================================================================================
\section{Proposed Task-specific denoising method}
\subsection{Theory}
\subsubsection{Problem formulation}
We propose the method in the context of reducing radiation dose in MPI SPECT, although the methodology can be applied in the context of reducing acquisition time. This is because reducing either dose or acquisition time eventually leads to a reduction in detected counts and the underlying objective of the proposed method is to denoise the low-count images. 

Consider a SPECT system imaging a tracer distribution (object) within the human body, described by a vector $f(\r)$,  where $\r \in \mathbb{R}^3$ denotes the 3-dimensional (3-D) coordinates, and yielding projection data, denoted by the M-D vector $\g$. Consider that the object and projection data lie in the Hilbert space $\mathbb{L}_2 (\mathbb{R}^3)$ and the M-D Euclidean space $\mathbb{E}^M$, respectively. Thus, the SPECT imaging system operator $\mathcal{H}$ maps object in $\mathbb{L}_2 (\mathbb{R}^3)$ to projection data in $\mathbb{E}^M$. The Poisson-distributed system-measurement noise is denoted by the M-D vector $\n$. The images are then reconstructed using the reconstruction operator, denoted by $\mathcal{R}$, yielding the reconstructed images, denoted by the \rchange{N}{$N_{3D}$}-D vector $\hat{\f}$. Thus,
\begin{align}
    \hat{\f} = \mathcal{R}\g = \mathcal{R}(\mathcal{H}\fobj + \n),\label{eq:f_hat}
\end{align}
where, without loss of generalization, we refer to the object as infinite-dimensional vector $\fobj$. From the reconstructed images, an observer performs the task of detecting perfusion defects. More precisely, the task is to classify the image into defect-absent ($H_0$) or defect-present ($H_1$) case. Denote the defect-absent object as $\fobj_b$ and the defect signal as $\fobj_s$. The two hypotheses for the defect-detection task are given by:
\begin{subequations}
\begin{align}
    &H_0: \hat{\f} = \mathcal{R}\g = \mathcal{R}(\mathcal{H}\fobj_b + \n), \label{eq:f_hat_h0}\\
    &H_1: \hat{\f} = \mathcal{R}\g = \mathcal{R}(\mathcal{H}(\fobj_b+\fobj_s) + \n).\label{eq:f_hat_h1}
\end{align}
\label{eq:f_hat_h0h1}%
\end{subequations}
In MPI SPECT, the perfusion-defect signal is a cold signal, so $\fobj_s$ is negative-valued.

In a low-dose protocol, the tracer uptake is lower compared to normal-dose protocols. Thus, the projection data, and the corresponding reconstructed images are noisier at low dose, impacting observer performance on the defect-detection task (Eq.~\eqref{eq:f_hat_h0h1}). Our goal is to design a technique to denoise low-dose images such that the denoised images yield improved performance on the defect-detection task.

\subsubsection{Proposed DL-based task-specific denoising method}
We consider the use of DL to design this denoising technique. Consider a deep network parameterized by the parameter vector $\Th$, denote the denoising operator by $\mathcal{D}_{\Th}$ and the predicted normal-dose image as $\hat{\f}_{ND}^{pred}$. The denoising operation can be mathematically expressed as follows:
\begin{align}
    \hat{\f}_{ND}^{pred} = \mathcal{D}_{\Th} (\hat{\f}_{{LD}}).
    \label{eq:denoising_eq}
\end{align}

To preserve the features that assist in the detection task while denoising, we propose a hybrid loss function for this deep network that consists of two terms. The first term penalizes the error associated with image fidelity between the actual and predicted normal-dose images. The second term penalizes the loss of features required to perform detection task in the predicted normal-dose images. Denote the fidelity loss term as $\mathcal{L}_{fid}(\Th)$ and the task-specific loss term as $\mathcal{L}_{task}(\Th)$. The hybrid loss function $\mathcal{L}(\Th)$ is given by:
\begin{align}
    \mathcal{L}(\Th) = \mathcal{L}_{fid}(\Th) + \lambda \mathcal{L}_{task}(\Th), \label{eq:hybrid_loss}
\end{align}
where $\lambda$ denotes a hyperparameter that controls the weights of these loss functions.

Denote the total number of training samples as $J$ and the $j^{th}$ sample of the low-dose image and normal-dose image by \rchange{$N$}{$N_{3D}$}-D vectors $\hat{\f}_{LD}^j$ and $\hat{\f}_{ND}^j$, respectively. Also, denote the normal-dose image predicted by the denoising network as $\hat{\f}_{ND}^{pred,j}$ when the low-dose image $\hat{\f}_{LD}^j$ is given as the input to the network. Thus,
\begin{align}
    \hat{\f}_{ND}^{pred,j} = \mathcal{D}_{\Th} (\hat{\f}_{{LD}}^j).
\end{align}

A typical choice to measure the fidelity between the actual and predicted normal-dose images, including in MPI SPECT, is the MSE between these images \cite{ramon2020improving,hwang2018improving}. Thus, we chose this distance measure as our fidelity-loss term. Consider that we have $J$ patient images in our training set. \rnew{Denote the number of voxels in each image slice by $N_{2D}$ and the number of slices as $Z$, so that $N_{3D}=N_{2D}×Z$. }Then, the fidelity-loss term is given by
\begin{align}
    \L_{\text{fid}}(\Th) = \frac{1}{J\rnew{N_{2D} Z}}\sum_{j=1}^{J}||\fhat_{ND}^j-\fhat_{ND}^{pred,j}||_2^2.
\end{align}

To obtain an expression for the task-specific loss term $\mathcal{L}_{task}(\Th)$ in Eq.~\eqref{eq:hybrid_loss}, we recognize that the detection task on MPI-SPECT images is performed by human observers. Thus, a mathematical term that preserves features used by human observers while performing detection tasks will intuitively assist in improving performance on the detection task. In this context, there is substantial literature on mathematical model observers that emulate human-observer performance \cite{abbey2001human,burgess1981efficiency,burgess1988visual,burgess1994statistically,loo1984comparison}. Further, multiple experiments in human vision have shown that the human visual system processes data using frequency-selective channels \cite{barrett2013foundations}. By processing the features extracted from these channels, referred to as anthropomorphic channels, studies have shown that model observers can mimic human-observer performance \cite{myers1987addition,frey2002application,abbey2001human}. Of most relevance to this paper, this has also been validated in studies with MPI SPECT on the task of detecting perfusion defects \cite{sankaran2002optimum,wollenweber1999comparison}. Thus, a denoising technique that preserves features extracted by these channels may assist with improving observer performance on detection tasks. 

Motivated by these studies, we design the task-specific loss term to preserve features that are derived by applying these anthropomorphic channels to the images. Typically, these channels are applied to the 2-D image slices. Thus, first, the profiles of the channels are centered on the defect location and the inner product of the channels and the to-be-processed 2-D image slices are computed to yield the feature value. Mathematically, denote \rold{the number of voxels in each image slice by $N_{2D}$, denote }an image slice by the $N_{2D}$-dimensional vector $\hat{\f}_{2D}$, denote the number of channels by $C$ and the $N_{2D}$-dimensional column vector corresponding to the $c^{th}$ channel by $\hat{\boldsymbol{u}}_c$. By concatenating the $C$ channel vectors, we obtain a $N_{2D}\times C$ \rchange{operator}{matrix} $\U$. Denote the shift operator that centers the channel profiles to the signal location by $\mathcal{S}$. \rnew{The shift operation on $\U$ can be represented by a multiplication of shift matrix $\boldsymbol{S}$ with the channel matrix $\U$. }The application of the \rnew{shifted }channel \rchange{operator}{matrix} on the centered image slice yields a $C$-dimensional vector, referred to as the channel vector and denoted by $\v$:
\begin{align}
    \v = (\boldsymbol{S}\U)^T\hat{\f}_{2D}.
\end{align}

The task-specific loss term $\L_{{task}}(\Th)$ penalizes the MSE between the channel vectors of the actual and predicted normal-dose image. \rold{Denote the number of slices in the images as $Z$, so that $N=N_{2D}×Z$. }To obtain the channel vector for the $j^{th}$ patient sample, we first perform acyclic 2-D shifting for each channel so that the center of the channel profile and centroid of the defect coincides. Since different patients will have defects at different locations, denote the shift \rchange{operator}{matrix} for the $j^{th}$ patient as \rchange{$\mathcal{S}^j$}{$\boldsymbol{S}^j$}. Also, denote the $s^{th}$ slice of the \rchange{image $\hat{\f}^j$ by $\hat{\f}_{2D,s}^j$}{normal-dose image $\hat{\f}_{ND}^j$ and the predicted normal-dose image $\hat{\f}_{ND}^{pred,j}$ by $\fhat_{ND,2D,s}^j$ and $\fhat_{ND,2D,s}^{pred,j}$, respectively}. The task-specific loss term $\L_{{task}}(\Th)$ is then given by:
\begin{align}
    \L_{{task}}(\Th) =& \frac{1}{J\rnew{C(s_2-s_1+1)}}\times\nonumber\\ 
    &\sum_{j=1}^{J}\sum_{s=s_1}^{s_2}||(\boldsymbol{S}^j\U)^{\text{T}}(\fhat_{ND,2D,s}^j-\fhat_{ND,2D,s}^{pred,j})||_2^2,\label{eq:task_loss}
\end{align}
where $s_1$ and $s_2$ denote the index of the start and end slices where the channels are applied, respectively.
\subsection{Implementation}
We developed an encoder-decoder architecture to minimize the loss function given by Eq.~\eqref{eq:hybrid_loss}. \rnew{The encoder-decoder architecture with multiple resolution levels was chosen motivated by the architectures previously proposed for denoising low-dose MPI-SPECT images \cite{yu2023ai,ramon2020improving}. }The schematic of the architecture is shown in Fig.~\ref{fig:arch}. The details of the network architecture are provided in the Supplementary material and in Rahman \etal \cite{rahman2023demist-supp-github}. The input and output to the network are the low-dose short-axis volume,  $\fhat_{LD}$  and the denoised (predicted normal-dose) short-axis volume, $\fhat_{ND}^{pred}$, respectively. The encoder extracts local spatial features from the low-dose image and generates a set of lower-dimensional latent features, which are used to reconstruct the denoised low-dose volume. Skip connections were used to add features learned in the encoder to the features generated by the decoder. Dropout was used to prevent overfitting. We trained the network by minimizing the hybrid loss in Eq.~\eqref{eq:hybrid_loss} using the ADAM algorithm \cite{kingma2014adam}.

\begin{figure}[h!]
\centering
\includegraphics[width = \linewidth]{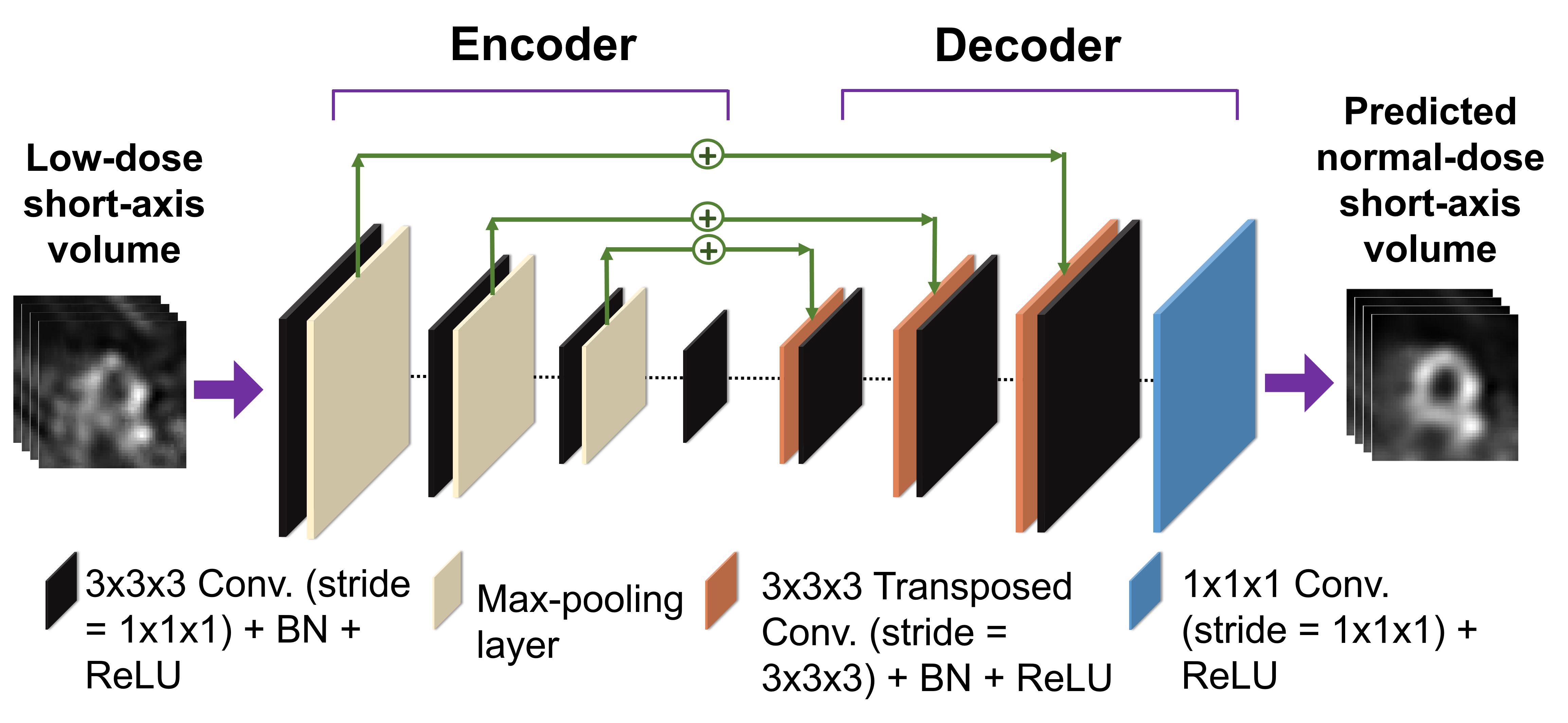}
\centering
\caption{A schematic with details of the encoder-decoder denoising network architecture. (BN = Batch normalization, ReLU = Rectified Linear Unit)}
\label{fig:arch}
\end{figure}

%In the training phase, to extract the channel vectors from defect-present images, as per Eq.~\eqref{eq:task_loss}, we shifted each channel profile in $\U$ to be centered to the defect centroid. As we describe later (Sec.~\ref{sec:eval_data_insertion}), we inserted synthetic defects in defect-absent images to obtain the defect-present images. Thus, the channel vectors for the corresponding defect-absent images were obtained by shifting the channel profiles in $\U$ to the centroid of the location where the synthetic defect was inserted. With these shifted channel profiles, we extracted the corresponding channel vectors from both predicted and normal-dose images and used these vectors to calculate the task-specific loss term (Eq.~\eqref{eq:task_loss}). More details of training the network are presented in Sec.~\ref{sec:net_train}.

%====================================================================================
% Evaluation
%====================================================================================
\section{Evalution}
We objectively evaluated the proposed method in an Institutional Review Board (IRB)-approved retrospective study conducted on clinical MPI-SPECT studies. We followed best practices for the evaluation of AI algorithms in nuclear medicine (RELAINCE guidelines) \cite{jha2022nuclear}.

\subsection{Data collection and curation}
We collected data from MPI studies ($N=4118$) conducted at clinical normal-dose level at Washington University School of Medicine. 
%The MPI-SPECT images in these studies were at clinical normal-dose level. 
\rnew{The clinical protocol was a one-day stress/rest protocol and the mean injected activity for the stress images was 10 mCi in patients weighing under 250 pounds and 12 mCi for those weighing over 250 pounds at normal-dose level. 
}
1295 MPI studies contained the \rnew{binned }SPECT projection data and CT images along with patient sex and anonymized clinical reports. The access to projection data allowed us to simulate the low-dose acquisition using binomial sampling \cite{juan2018investigation}\rnew{, which preserved the Poisson noise process in the low-dose projections \cite{barrett2013foundations}. For Binomial sampling, we used MATLAB’s default Mersenne Twister algorithm for pseudo-random number generation}. We considered low-dose levels of 25\%, 12.5\% and 6.25\%. \rnew{In generating the low-dose levels, we assumed that the fractional myocardial tracer uptake is linearly related to the injected dose. Thus, the count levels in myocardial wall were 25\%, 12.5\% and 6.25\% of normal-dose count level. }At these low-dose levels, the performance on detection task is significantly different than normal-dose images \cite{yu2023ai}, and task performance is dominated by system noise compared to anatomic variability in patient populations \cite{he2010investigation}. Thus, choosing these dose levels provided a regime to study the efficacy of the proposed method in improving task performance over low-dose images. 

 For training and evaluation of the proposed method, both the knowledge of presence of defect and the defect centroid were needed. Although presence of defect could be read from the clinical reports, findings in these reports often suffer from reader variability. Moreover, the defect centroid is typically unavailable. To address this issue, we only used the normal (defect-absent) MPI studies ($N = 795$) and inserted synthetic defects using a defect-insertion approach described later (Sec.~\ref{sec:eval_data_insertion}) to create the defect-\rchange{absent}{present} images. For defect insertion, segmentation of the left ventricle (LV) wall was needed, but this wall could not be segmented reliably for some cases. Also, in some other cases, the images contained artifactual (apparent) defects. In clinical practice, these artifactual defects are typically ruled out using other patient data, such as the rest scans, polar maps, and projection scans. However, in our observer study, only the stress images are used for the detection task. Thus, we excluded these two sets of cases ($N=457$) and only used the remaining normal cases ($N=338$).  
 %The data for these 338 normal cases were acquired across four SPECT/CT scanners: GE TANDEM APOLLO ($N=157$), GE TANDEM DISCOVERY 670 ($N=129$), GE TANDEM 870 CZT ($N=4$) and GE TANDEM DISCOVERY 670 PRO ($N=48$) with a rest/stress protocol. The data-collection process is illustrated in Fig.~\ref{fig:pat_dist}.
% The data for these 338 normal cases were acquired across multiple SPECT/CT scanners including GE DISCOVERY 670 PRO and GE 670 CZT with a rest/stress protocol. The data-collection process is illustrated in Fig.~\ref{fig:pat_dist}.
%More details on the acquisition and reconstruction parameters of the SPECT/CT scanners are provided in the Supplementary material as well as in Rahman \etal \cite{rahman2023demist-supp-github}.
The datasets were from two scanners, namely the GE Discovery 670 Pro NaI and the GE Discover\rnew{y} 670 CZT. These two systems have different detectors, namely NaI and CZT, each of which have different energy and position resolutions (as \rchange{indicated}{listed} in the supplementary material and in \cite{rahman2023demist-supp-github}). 
%Thus, we studied the performance of the method using data from two different SPECT system configurations. 

\begin{figure}[h!]
\centering
\includegraphics[width = \linewidth]{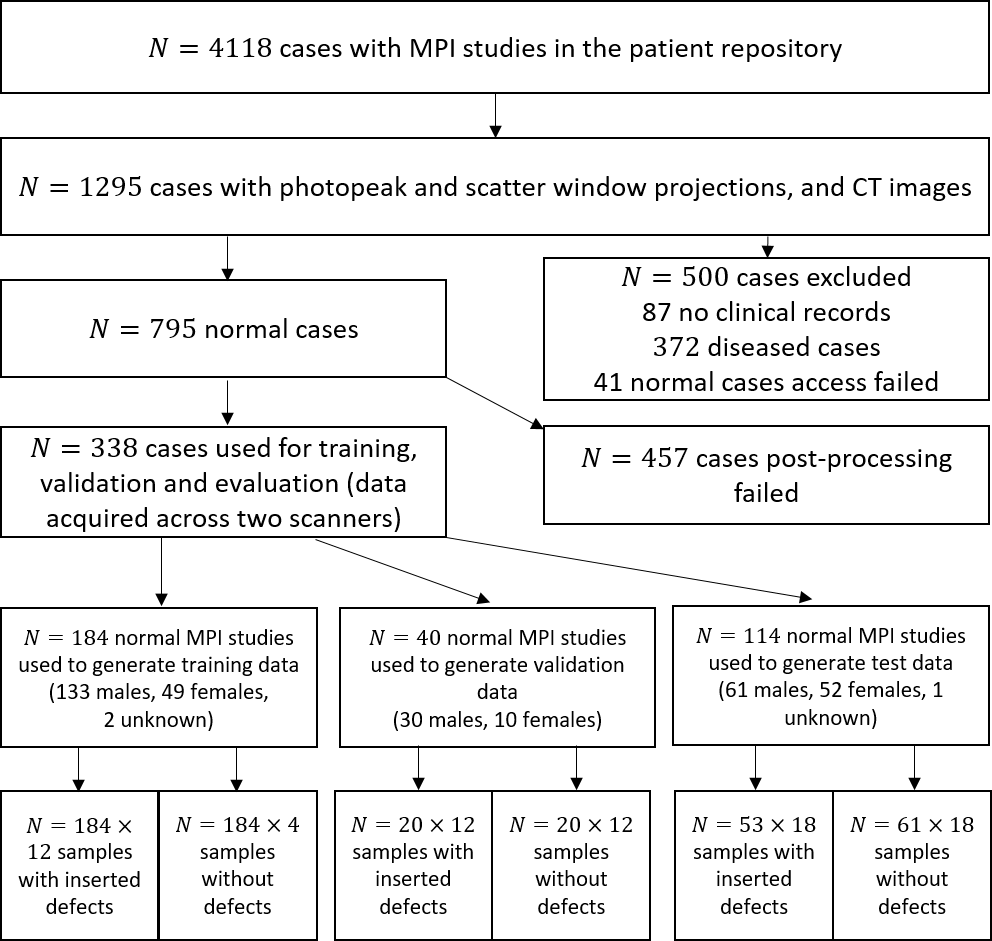}
\caption{Patient data collection from MPI studies and their distribution in various stages of data curation.}
\label{fig:pat_dist}
\end{figure}

\subsubsection{Defect insertion approach}\label{sec:eval_data_insertion}
To insert the defect, we first segmented the LV wall using the reoriented short-axis normal-dose image using SEGMENT software \cite{heiberg2010design,soneson2009improved}. From the centroid of the LV wall, a 2-D cone region with a specific extent was located. For anterior-wall defect, the cone region was between 80$^{\circ}$ and $(80 - \theta)^\circ$ where $\theta$ denotes the defect extent  and was assigned values of 30$^{\circ}$ and 60$^{\circ}$. For determining the angles, the x-axis was assumed to be along the rows of the reoriented image and the origin was the centroid of the LV wall. For inferior-wall defect, the cone region was between -80$^{\circ}$ and $(-80 + \theta)^\circ$. In the slice containing the LV centroid, the LV wall that lies inside this cone region was considered as the defect mask. The same cone region was used in adjacent slices to create the 3D defect. A 42 mm defect in the long-axis direction (apex-to-base direction) was considered. We used the mean LV uptake as reference to define defects with specific severities. The defect signal, with specific severity and extent, was then subtracted from the reconstructed image of the defect-absent case to create an initial defect-present image. Next, to create the hybrid dataset with inserted defects, we employed a strategy similar to that proposed by Narayanan \etal \cite{narayanan2001optimization}. Briefly, we used SIMIND, a well-validated Monte-Carlo simulation software \cite{ljungberg1989monte,toossi2010simind} to generate the intermediate projection data corresponding to the defect-absent image and the initial defect-present image. These intermediate projection data were then used to calculate a scale factor. The clinical projection data from defect-absent cases were scaled using this scale factor to create the final defect-present projection data.

\subsubsection{Reconstruction and post-processing}
We used a clinical reconstruction protocol based on the ordered subset expectation maximization (OSEM) algorithm \rnew{implemented with CASTOR \cite{merlin2018castor} }to reconstruct the normal-dose and low-dose images. The reconstruction compensated for attenuation and collimator-detector response. \rnew{The number of subsets and the iterations in the OSEM algorithm was selected based on the protocol used in the clinic. }3-D Butterworth filtering with filter order of 5 and cutoff frequency of 0.44 \rchange{cm$^{-1}$}{cycles/cm} was applied to the images, which were then reoriented to the short axis using linear interpolation. From this reoriented image, we extracted a $48\times 48\times 48$ volume where the center of the volume coincided with the center of LV. For better dynamic range, we set the range of the pixel values to [0, $x_{LV}$] where  $x_{LV}$ is the maximum value inside the LV wall.

\subsection{Network training} \label{sec:net_train}
The training set consisted of 2944 cases. These were obtained from 184 normal MPI studies. A total of 12 synthetic defect types were generated for each normal study, where the defect types were defined in terms of their extent, severity, and position in the LV wall. The defects were inserted in the anterior and inferior walls, had extents of 30$^{\circ}$ and 60$^{\circ}$, and severities of 10\%, 17.5\% and 25\%. We inserted these 12 defect types in each of the 184 normal studies to generate the defect-present population ($N= 2208$). The defect-absent population was obtained by replicating the 184 normal studies a total of \rchange{six}{four} times, corresponding to the \rchange{six}{four} different defect extents and locations. Thus, the defect-absent population consisted of $N=184\times 4=736$ samples. These two populations, totaling $N = 2944$ cases, were used to train the network. 

In the training phase, to extract the channel vectors from defect-present images, as per Eq.~\eqref{eq:task_loss}, we shifted each channel profile in $\U$ to be centered to the defect centroid. 
%As we describe later (Sec.~\ref{sec:eval_data_insertion}), we inserted synthetic defects in defect-absent images to obtain the defect-present images. 
The channel vectors for the corresponding defect-absent images were obtained by shifting the channel profiles in $\U$ to the centroid of the location where the synthetic defect was inserted. With these shifted channel profiles, we extracted the corresponding channel vectors from both predicted and normal-dose images and used these vectors to calculate the task-specific loss term (Eq.~\eqref{eq:task_loss}). 
%More details of training the network are presented in Sec.~\ref{sec:net_train}.
We performed a four-fold cross-validation to optimize the network. The training was performed on an NVIDIA TESLA V100 GPU with 32 GB of RAM. We trained separate networks for each dose level and a range of $\lambda$ values.
To select the optimized $\lambda$ value for each dose level, we used a separate validation set obtained from 40 normal cases. Using the same strategy as for the training set, 20 of these cases were used to create the defect-present population of $20\times 12=240$ samples. For a specific low-dose level, we denoised the images in the validation set using pre-trained networks corresponding to different $\lambda$ values. Using observer studies, as will be described in Sec.~\ref{sec:eval_testing_proc}, for each dose level, the value of $\lambda$ that maximized performance on the detection task was selected as the optimal $\lambda$. 

\subsection{Testing procedure}\label{sec:eval_testing_proc}
The test set consisted of $N = 2052$ cases. These were generated using $N=114$ normal MPI studies. Of these, 61 normal studies were used as the defect-absent population. To create the defect-present population, synthetic defects were inserted in the 53 normal studies. In addition to the 12 defect types, we also introduced six new defects with 45$^\circ$ extent to create out-of-distribution defect types in the test set. These new defects had severities and locations as the usual defects. Therefore, the test set consisted of 18 types of defects.

We evaluated the performance of the proposed method on the clinical task of detecting perfusion defects and using task-agnostic fidelity-based figures of merit. 
%Performance was compared to low-dose protocols and to a commonly used DL-based denoising method \cite{ramon2020improving} that was trained with a loss function that used only the fidelity term (setting $\lambda = 0$ in Eq.~\eqref{eq:hybrid_loss}). We refer to this method as task-agnostic DL-based denoising (TADL).
Performance was compared to low-dose images that were not denoised. We refer to this as the low-dose protocol. To assess the impact of using our task-specific denoising strategy, we also compared performance to images that were denoised using a commonly used DL-based denoising method \cite{ramon2020improving} that was trained with a loss function that used only the fidelity term (setting $\lambda = 0$ in Eq.~\eqref{eq:hybrid_loss}). We refer to this method as the task-agnostic DL-based denoising (TADL) method.\rnew{ Comparing DEMIST with TADL method allowed assessing the impact of incorporating the task-specific term into the loss function on observer performance.}

To objectively evaluate the proposed method on the task of detecting perfusion defects, we considered an anthropomorphic channelized Hotelling observer (CHO) \cite{myers1987addition} as a surrogate for the human observer. For clinical application, ideally the performance of the proposed method on the defect-detection task should be evaluated using human-observer studies by trained radiologists. However, such studies are time-consuming, expensive, tedious, and may be inappropriate at early stage of method development. 
%Mathematical model observers have been developed to emulate the performance of human observers, of which the channelized Hotelling Observer (CHO) is widely used \cite{myers1987addition}. 
To address this challenge, model observers such as the CHO \cite{myers1987addition} have been developed. 
Most importantly, CHOs with rotationally symmetric frequency channels have been validated to emulate human-observer performance on the task of detecting location-known perfusion defects in MPI SPECT \cite{sankaran2002optimum,wollenweber1999comparison}. Thus, we used the CHO with these channels as our observer.\rnew{ We follow the same procedure as in \cite{wollenweber1999comparison} to define the rotationally symmetric frequency channels. Briefly, the start frequency and bandwidth of first channel is 0.1838 cycles/cm. The subsequent channels were adjacent to the previous one and had double the start frequency and bandwidth as the previous one.} 

We selected the 2-D short-axis slice and two adjacent slices from each MPI-SPECT image that contained the defect centroid for conducting the observer studies. From the centroid-containing slice, \rnew{consistent with previous studies \cite{frey2002application}, }we extracted a $32\times 32$ region such that the defect centroid was at the center of the extracted region. This same 2-D region was also extracted from the two adjacent slices. Pixels values of each extracted region were mapped to the range [0,255]. We then applied anthropomorphic rotationally symmetric frequency channels to each slice to compute the channel vectors. The channel vectors of defect-present and defect-absent populations were used to learn the template of the CHO using a leave-one-out approach. Following that, the test statistics were computed and used to perform the ROC analysis. Stratified \rchange{analysis}{analyses} based on sex, defect severity \rchange{and }{,}defect extent \rnew{and scanner type }were also performed. 

\subsection{Figures of merit}
ROC analysis was performed on the test statistics derived with the CHO using the pROC package in R \cite{pROCSoft}. The area under the empirical ROC curve (AUC) was used as the figure of merit. Confidence intervals were calculated using Delong’s method \cite{delong1988comparing}, which accounts for variability across cases. The AUC values were computed for the normal-dose and low-dose images and those denoised with DEMIST and TADL. To test the statistical significance of difference in AUC values between two methods, we used Delong’s test as implemented within the pROC package \cite{pROCSoft}. To account for multiple hypothesis testing (DEMIST vs. LD, DEMIST vs. TADL and TADL vs LD), we used Bonferroni correction \cite{bland1995multiple}. 
A corrected $p$ value $< 0.05$ was used to infer a statistically significant difference.
For evaluation based on image fidelity, we considered two widely used fidelity-based figures of merit: RMSE and SSIM.

%====================================================================================
% Result
%====================================================================================
\section{RESULTS}

\subsection{Evaluation on the task of perfusion defect detection}
\begin{figure}[h!]
\centering
\includegraphics[width = 0.9\linewidth]{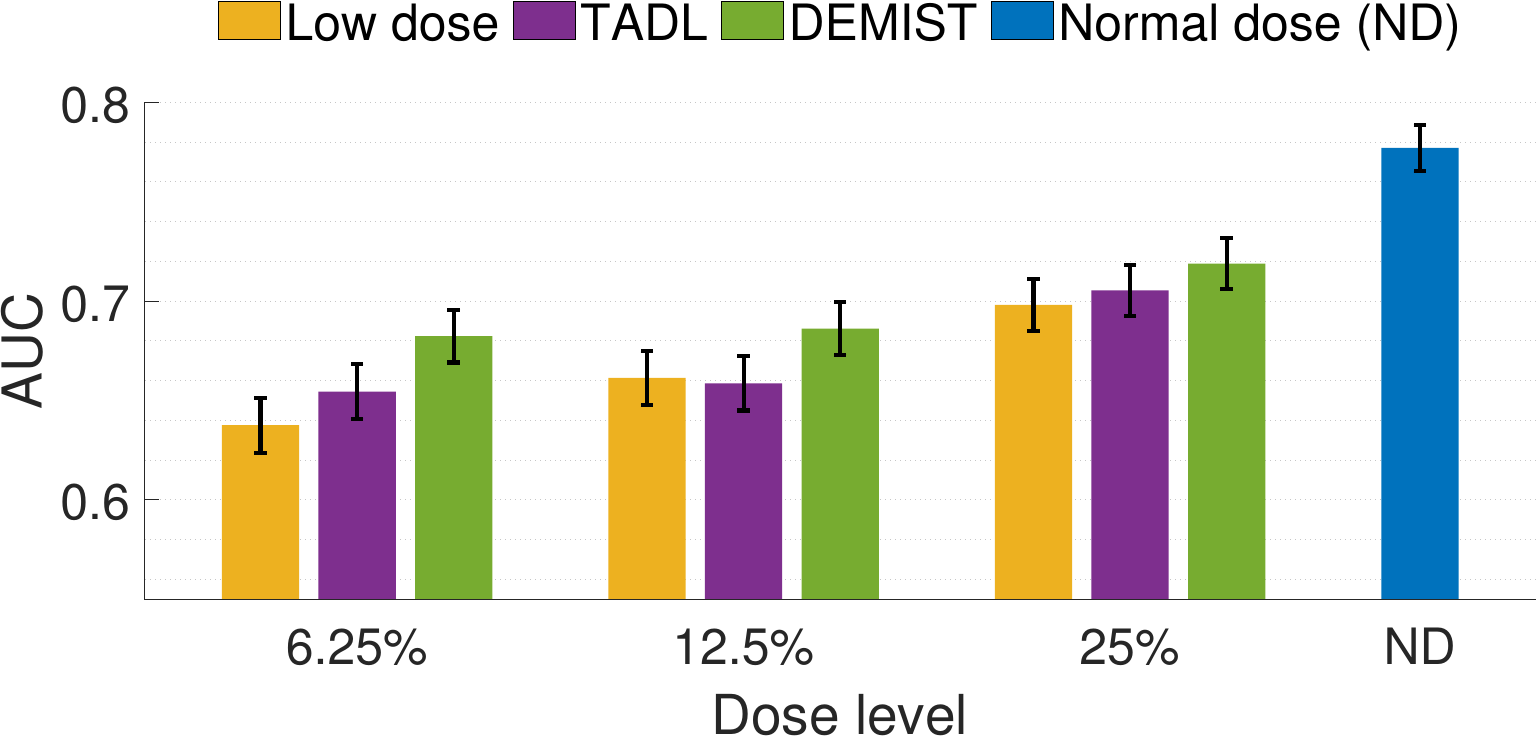}
\caption{AUC values obtained for the normal and low-dose images, and the images denoised using the DEMIST and TADL approaches at various dose levels with CHO. Error bars denote 95\% confidence intervals.
}
\label{fig:auc}
\end{figure}

Fig.~\ref{fig:auc} shows the AUC values obtained with the low-dose protocol, DEMIST, and TADL methods at all the considered low-dose levels, and with the normal-dose protocol. At all dose levels, DEMIST significantly outperformed low-dose protocol as well as the TADL method. 
The $p$-values of all the statistical tests presented in these results are included in the Supplementary material and in Rahman \etal \cite{rahman2023demist-supp-github}.
We also note that, in most cases, the AUC values obtained with the TADL method and with the low-dose protocol were similar. We do note that the proposed method yields inferior performance on detection task compared to normal-dose protocol, an observation that we will discuss in the Discussions section. 
%The $p$-values of all statistical tests are included in the Supplementary material and in Rahman \etal \cite{rahman2023demist-supp-github}.

\begin{figure*}[h!]
\centering
\includegraphics[width = 0.9\linewidth]{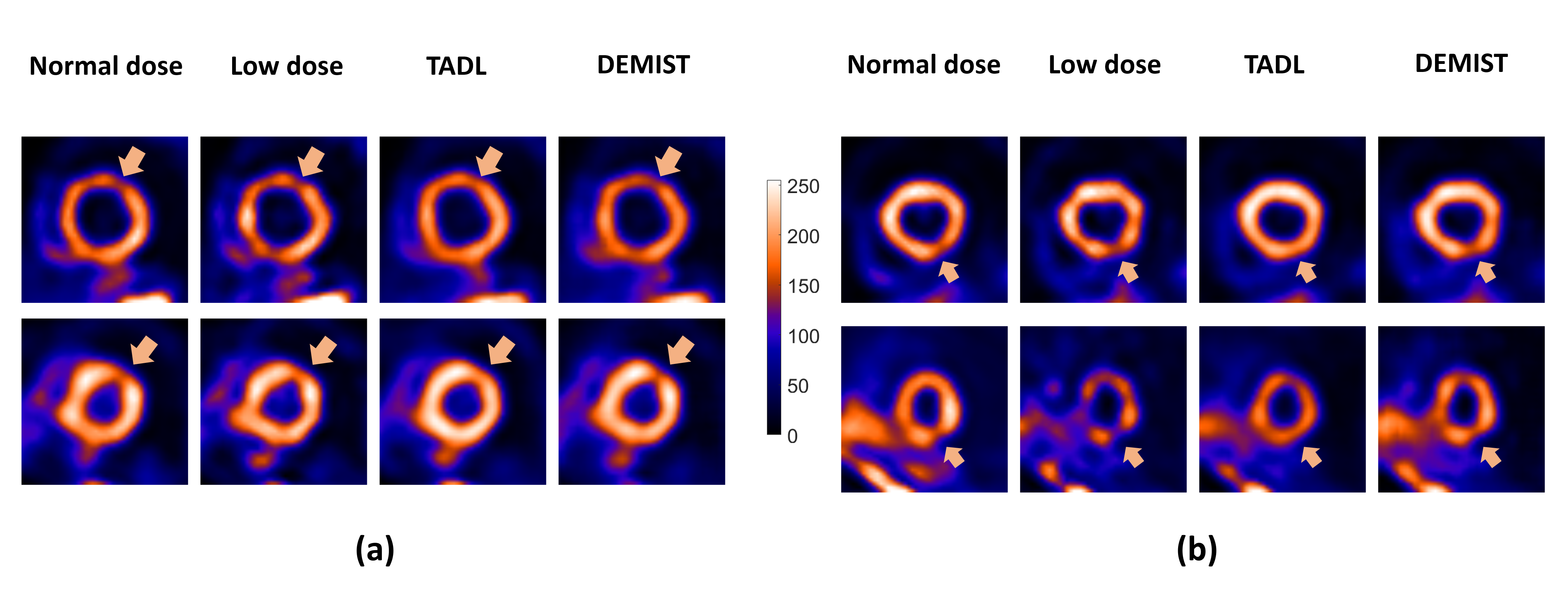}
\caption{Four representative tests cases qualitatively showing the performance of TADL method and proposed DEMIST method. For all cases, the low-dose level was set to 12.5\%. In (a) and (b), defects were in anterior and inferior wall, respectively. For all four cases, the defects had an extent of 30$^\circ$ and severity of 25\%. First, we note that the background appears less noisy compared to low-dose images with both TADL and DEMIST. The defect tends to be become less detectable with the TADL (no task-specific loss term). However, the defect was visually clearer with the proposed DEMIST method.}
\label{fig:lambda_quality}
\end{figure*}

Fig.~\ref{fig:lambda_quality} qualitatively shows the impact of the DEMIST and TADL methods on four representative cases. We observe in these cases that with the TADL method, even though the background looks less noisy compared to low-dose protocol, the defect tends to wash out.
This observation is consistent with the findings reported in previous studies \cite{yu2023ai,ongie2022optimizing}.
In contrast, with DEMIST, the defect is visibly clearer even as the background looks less noisy compared to low-dose protocol. These representative cases provide an intuitive explanation for the improved performance of the DEMIST method.

\begin{figure}[h!]
\centering
\includegraphics[width = 0.9\linewidth]{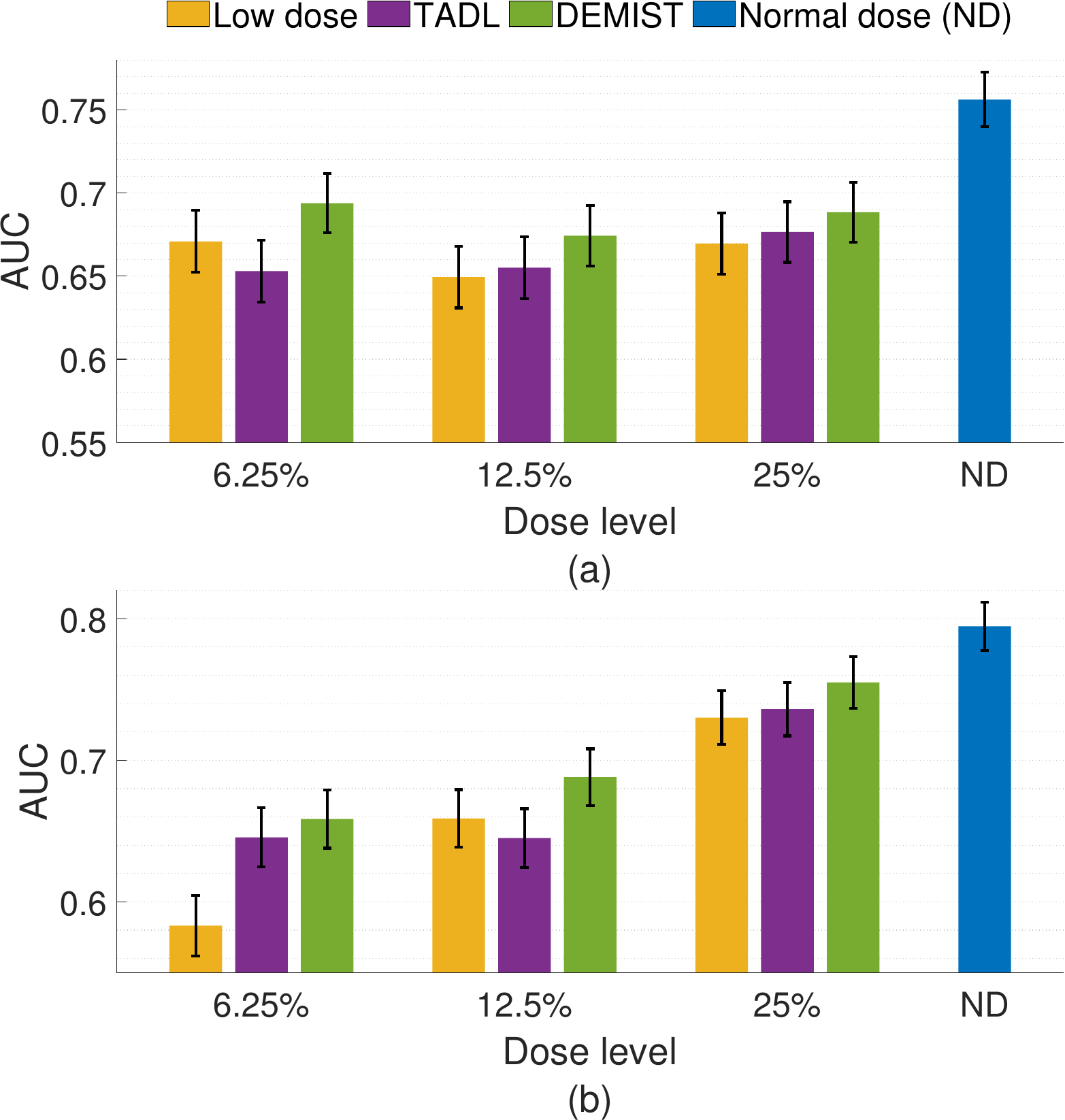}
\caption{AUC values obtained for the different approaches and at various dose levels with (a) male and (b) female patients using CHO. Error bars denote 95\% confidence intervals.
}
\label{fig:auc_mf}
\end{figure}

Figs.~\ref{fig:auc_mf}a and \ref{fig:auc_mf}b show the AUC values obtained with male and female populations, respectively. We observed that, for both sexes, the proposed method yielded a significant improvement in performance on the detection task at all dose levels compared to low-dose protocol. Moreover, in 5 out of 6 settings (3 dose levels $\times$ 2 sexes), DEMIST yielded significant improvement in detection-task performance compared to TADL method. Further, again, the TADL method generally did not improve (and in some cases degraded) performance compared to the low-dose protocol. 

\iffalse
\begin{figure}[h!]
\centering
\includegraphics[width = 0.9\linewidth]{images/auc_male_cho.eps}
\caption{AUC values for male patients obtained for the different approaches and at various dose levels using CHO. Error bars denote 95\% confidence intervals.
}
\label{fig:auc_male}
\end{figure}

\begin{figure}[h!]
\centering
\includegraphics[width = 0.9\linewidth]{images/auc_female_cho.eps}
\caption{AUC values obtained for the different approaches and at various dose levels with female patients using CHO. Error bars denote 95\% confidence intervals.
}
\label{fig:auc_female}
\end{figure}
\fi

\begin{figure*}[h!]
\centering
\includegraphics[width = 0.8\linewidth]{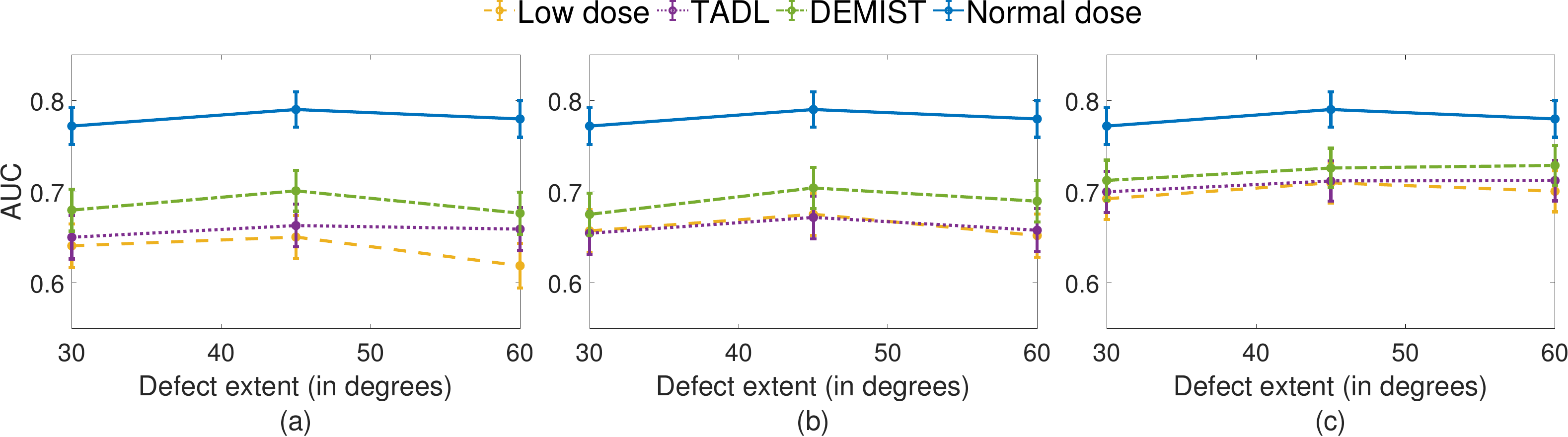}
\caption{AUC values obtained using CHO for the various approaches as a function of different defect extents with (a) 6.25\%, (b) 12.5\% and (c) 25\% dose levels. Error bars denote 95\% confidence intervals.
}
\label{fig:auc_def_ext_cho}
\end{figure*}

\begin{figure*}[h!]
\centering
\includegraphics[width = 0.8\linewidth]{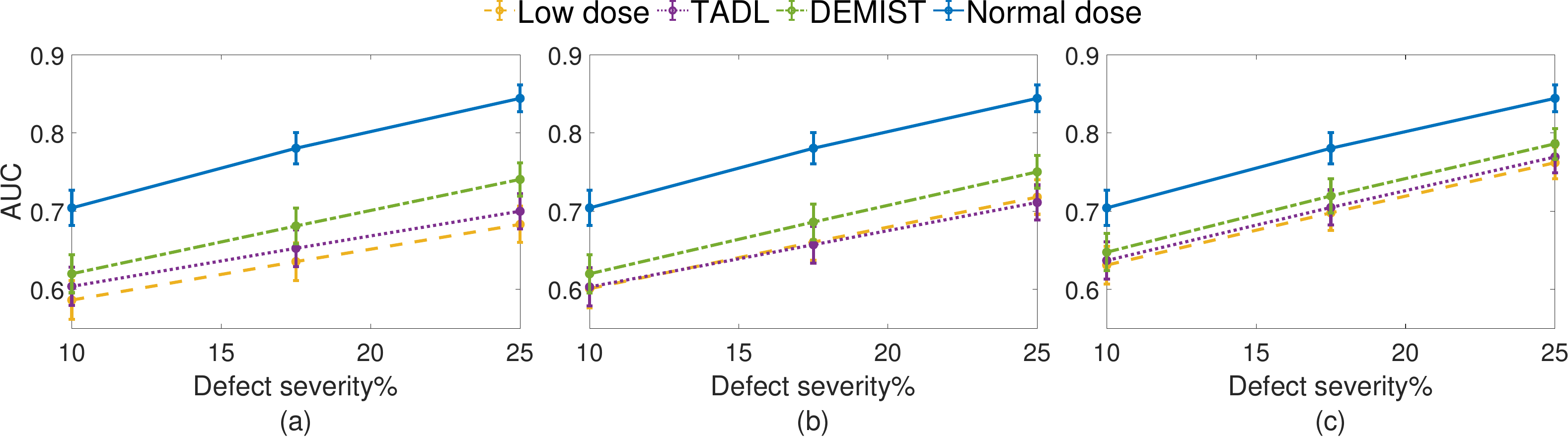}
\caption{AUC values obtained using CHO for the various approaches as a function of different defect severities with (a) 6.25\%, (b) 12.5\% and (c) 25\% dose levels. Error bars denote 95\% confidence intervals.
}
\label{fig:auc_def_sev_cho}
\end{figure*}

Figs.~\ref{fig:auc_def_ext_cho} and \ref{fig:auc_def_sev_cho} show the AUC values as a function of defect extent and severity at different dose levels, respectively. We observe that, at all dose levels, the DEMIST method significantly improved observer performance for all considered defect extents and severity compared to low-dose protocol. Moreover, the DEMIST method significantly improved observer performance compared to TADL method in 15 out of 18 settings (3 dose levels $\times$ 6 defect types). Again, the TADL method was generally observed to not improve performance compared to low-dose protocol.

\begin{figure*}[h!]
\centering
\includegraphics[width = 0.9\linewidth]{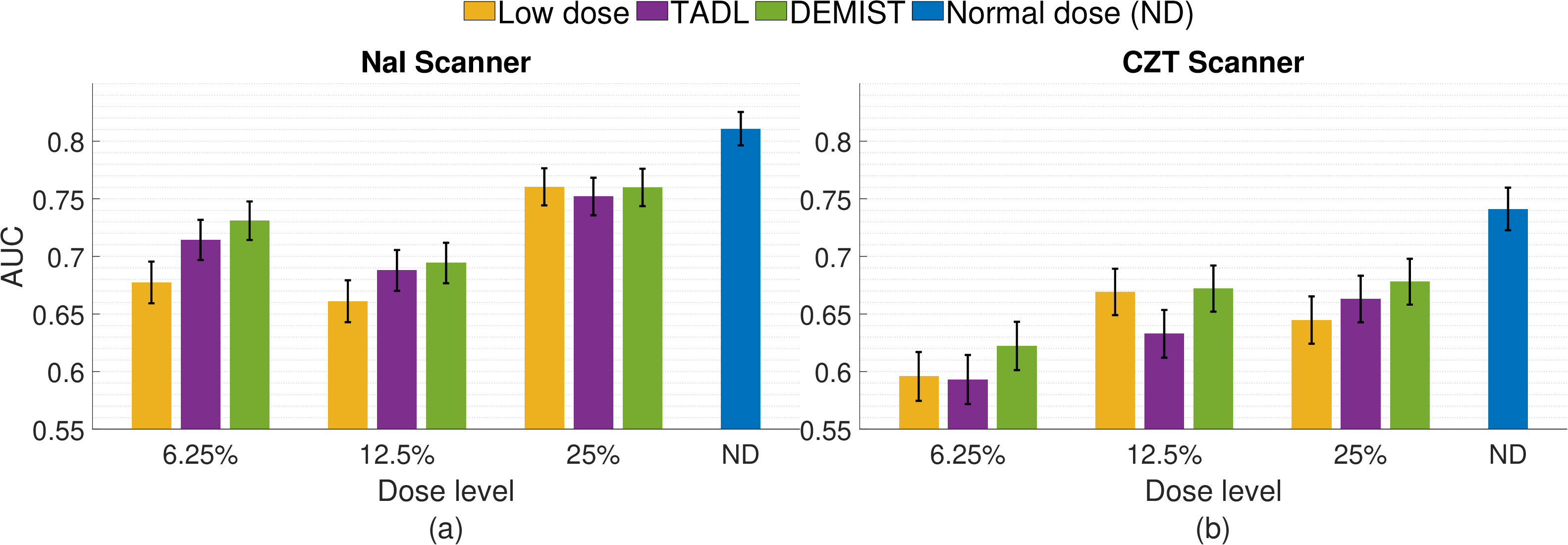}
\caption{AUC values obtained for the considered approaches and at different dose levels with data from the (a) NaI and (b) CZT scanners using CHO. Error bars denote 95\% confidence intervals.
}
\label{fig:auc_scanner}
\end{figure*}

\rnew{Fig. \ref{fig:auc_scanner} shows the AUC values obtained for stratified analysis based on scanner models. In our study, data were collected across two scanners, namely “GE Discovery NM/CT 670 Pro NaI” and “GE Discovery NM/CT 670 Pro CZT”. For conciseness, we refer to these two scanners as NaI and CZT scanner, respectively. We observe from Fig.~\ref{fig:auc_scanner} that the DEMIST method significantly outperformed low-dose protocol in 4 out of 6 settings (3 dose levels $\times$ 2 scanners) and  the TADL method in 4 out of 6 settings. We also observed that the performance of the TADL method deteriorated by comparison with the low-dose protocol in some settings. These findings demonstrate the robustness of the proposed method across different scanner types.
}

\subsection{Evaluation based on fidelity-based figures of merit}
The SSIM and RMSE metrics are presented in Table 1 for the proposed DEMIST method, TADL method and low-dose protocol. We observed that DEMIST yielded improved performance compared to low-dose protocol. Moreover, in general, both the proposed DEMIST method and the TADL method yielded very similar RMSE and SSIM values.

\begin{table}[h!]
\centering
\caption{RMSE AND SSIM METRIC FOR DIFFERENT METHOD AT DIFFERENT DOSE LEVELS. }
\label{tab:rmse_ssim}
\begin{tabular}{|c|c|c|c|c|}
\hline
\begin{tabular}[c]{@{}c@{}}Dose \\ level\end{tabular} &
  Metric &
  Low dose &
  TADL &
  DEMIST \\ \hline
\multirow{2}{*}{6.25\%} & RMSE & 6.87 & 5.00 & 5.58 \\ \cline{2-5} 
                        & SSIM & 0.77 & 0.85 & 0.84 \\ \hline
\multirow{2}{*}{12.5\%} & RMSE & 4.81 & 4.10 & 4.01 \\ \cline{2-5} 
                        & SSIM & 0.86 & 0.89 & 0.89 \\ \hline
\multirow{2}{*}{25\%}   & RMSE & 3.16 & 2.94 & 2.94 \\ \cline{2-5} 
                        & SSIM & 0.93 & 0.93 & 0.94 \\ \hline
\end{tabular}
\end{table}
%====================================================================================
% Discussion
%====================================================================================

\section{DISCUSSION}
In this work, we proposed a method to denoise low-dose MPI-SPECT images while preserving features that assist in performing detection task by incorporating a task-specific loss term. \rnew{We then evaluated our method on the clinical task of detecting perfusion defects in MPI-SPECT using a retrospective clinical study. }
The result in Fig.~\ref{fig:auc} shows that applying this method resulted in significantly improved defect-detection performance over just using low-dose images \rchange{or denoising the}{, as well as} low-dose images \rchange{with a commonly used task-agnostic DL}{denoised using the TADL} method. 
Similar results were observed when the method was analyzed on test data with specific sub-groups based on sex and defect types (Figs.~\ref{fig:auc_mf}-\ref{fig:auc_def_sev_cho}). These results provide evidence that incorporating this task-specific loss term can significantly improve observer performance beyond low-dose images and using a commonly used task-agnostic DL-based denoising method consistently across a range of defect characteristics.
\rnew{ To the best of our knowledge, this is the first time that a DL-based denoising method for MPI SPECT has shown improved performance on the task of detecting perfusion defects in an anthropomorphic model-observer study.} 

%In the representative qualitative results (Fig.~\ref{fig:lambda_quality}), we observed that the TADL method, which only incorporates fidelity-based loss, tends to wash out the defect. This observation is consistent with the findings reported in previous studies \cite{yu2023ai,ongie2022optimizing}.  For the cases shown in Fig.~\ref{fig:lambda_quality}, we observed that by incorporating the task-specific loss term, the defect becomes more visually detectable. Thus, these results provide visual intuition for the improved performance of DEMIST on the detection task.

\rchange{Fig.~\ref{fig:lambda_quality} provided visual insights on}{To mathematically interpret} the improved performance of the \rchange{proposed}{DEMIST} method\rold{. To mathematically interpret this improved performance}, we conducted an analysis similar to Yu \etal \cite{yu2023ai}. More specifically, we analyzed the effect of denoising on the first and second-order \rchange{statistic}{statistics} of \rnew{the }channel vector\rnew{s} of the test set for both DEMIST and TADL. The analysis was performed for each defect type separately. Denote the mean difference channel vector between defect-present and defect-absent cases as $\Delta \bar{\v}$ and the channel-vector covariance matrix as $\Kb_{\v}$. The signal-to-noise ratio (SNR) of the CHO is given by:
\begin{align}
    \text{SNR}^2 = \Delta \bar{\v}^{\text{T}} \Kb_{\v}^{-1} \Delta \bar{\v}.
    \label{eq:snr_cho}
\end{align}
If the test statistics of defect-\rchange{present}{absent} and defect-present cases are normally distributed, AUC and SNR of the observer are monotonically related \cite{barrett2013foundations} and thus, the analysis of observer SNR yields insights on detection-task performance. 

\begin{figure*}[h!]
\centering
\includegraphics[width = \linewidth]{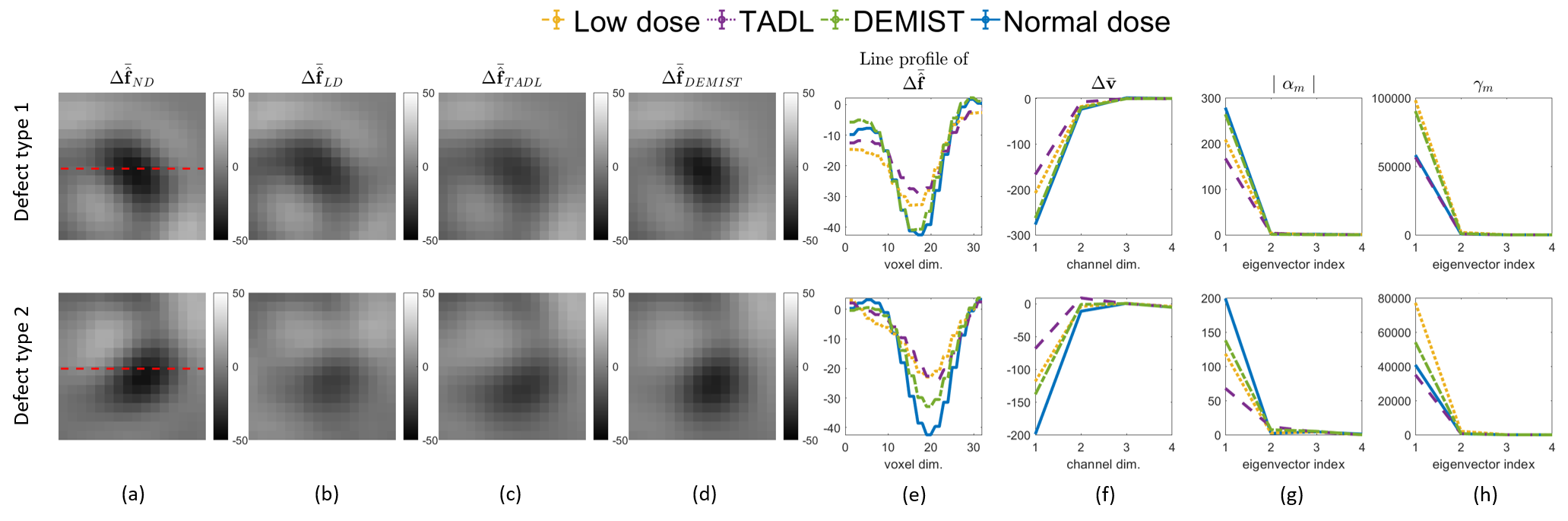}
\caption{Mean difference reconstructed short-axis image between defect-absent and defect-present hypotheses for (a) normal-dose, (b) low-dose, (c) TADL and (d) DEMIST. The images in (a-d) are windowed to a region centered to the defect centroid. (e) Line profile of the mean difference reconstructed image of (a-d). \rnew{The red dashed lines in (a) represent the lines along which the profiles are drawn. }(f) Mean difference channel vector 
$\Delta \bar{\v}$ between defect-absent and defect-present hypotheses for various approaches. (g) Absolute value of coefficient $\alpha_m$ and (h) eigenvalue spectra of noise covariance matrix. Low-dose level was set to 6.25\%. (Defect type 1: 60$^\circ$ extent, 25\% severity and anterior wall defect. Defect type 2: 45$^\circ$ extent, 25\% severity and inferior wall defect.)}
\label{fig:svd_analysis}
\end{figure*}

Consider that the reconstructed images have been reoriented and windowed with defect centroid at the center.
Denote the mean difference reconstructed image between defect-present and defect-absent cases by $\Delta\bar{\fhat}$. Thus, $\Delta \bar{\v}=\U^\text{T} \Delta\bar{\fhat}$. As per Eq.~\eqref{eq:snr_cho}, both the mean difference of the channel vector $\Delta \bar{\v}$ and covariance matrix $\Kb_{\v}$ affect observer performance. Eigenanalysis of the covariance matrix provides a mechanism to analyze the combined effect of these two terms \cite{yu2023ai} on the observer SNR. Denote the $m^{th}$ eigenvector and eigenvalue of $\Kb_{\v}$ by $\boldsymbol{u}_m$ and $\gamma_m$, respectively. We can express $\Delta \bar{\v}$ in terms of these eigenvectors as follows:
\begin{align}
    \Delta \bar{\v}=\sum_{m=1}^{C}\alpha_m \boldsymbol{u}_m, \label{eq:delv_decomp}
\end{align}
where the coefficient $\alpha_m=\boldsymbol{u}_m^\text{T} \Delta \bar{\v}$. Further, the SNR of the CHO is given by \cite{yu2023ai}
\begin{align}
    \text{SNR}^2 = \sum_{m=1}^{C} \dfrac{\alpha_m^2}{\gamma_m}.
\end{align}
Thus, assessing the impact of denoising on $\alpha_m$ and $\gamma_m$ provides an interpretable approach to evaluate the effect of denoising on observer performance. Fig.~\ref{fig:svd_analysis} shows this analysis for two defect types with 6.25\% dose level. We first plotted the mean difference reconstructed image $\Delta\bar{\fhat}$ and mean difference channel vector ($\Delta \bar{\v}$) between defect-present and defect-absent cases (Figs.~\ref{fig:svd_analysis}a-\ref{fig:svd_analysis}f). We observe that the DEMIST method preserved the mean difference originally present in the normal-dose image for this defect type. However, as in Yu \etal \cite{yu2023ai}, we observed that the TADL method reduced this mean difference, negatively impacting observer performance. Figs.~\ref{fig:svd_analysis}g-\ref{fig:svd_analysis}h show the values of $\alpha_m$ and $\gamma_m$ as a function of $m$. We observed that $\gamma_m$ reduced for both DEMIST and TADL method compared to low-dose images, which would positively impact observer performance. However, with the TADL method, the values of $\alpha_m$ were lower compared to low-dose images, which leads to limited observer performance on detection task. 
%This provides an interpretation for the limited performance of the TADL technique on the defect-detection task. 
In contrast, with the DEMIST method, the values of $\alpha_m$ do not reduce (and in some cases increase) compared to low-dose images, resulting in an overall improvement in performance on the defect-detection task.

\begin{figure*}[h!]
\centering
\includegraphics[width = 0.9\linewidth]{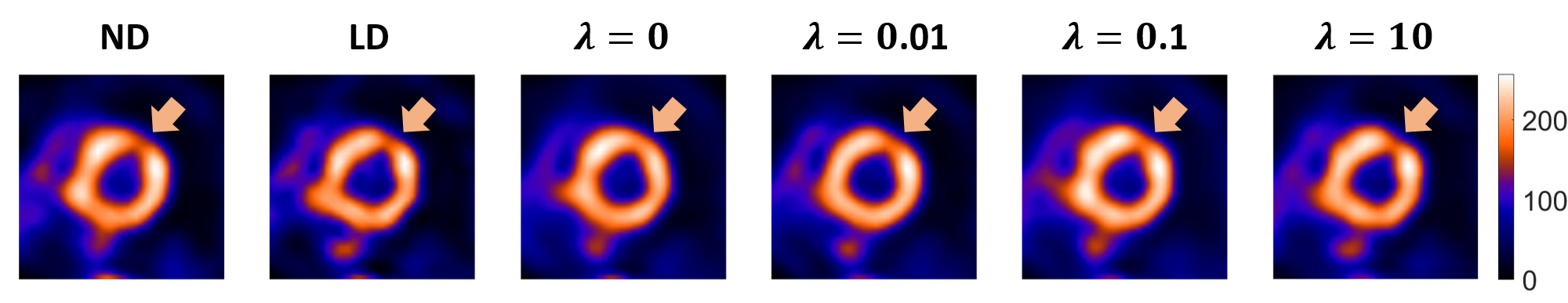}
\caption{From left, normal-dose (ND) image, low-dose (LD) image acquired at 12.5\% dose level and images denoised with proposed method with varying  $\lambda$ as indicated on top of the image. Increasing $\lambda$ results in recovery of defect visibility. However, the increase in $\lambda$ also results in a decrease in background smoothness.}
\label{fig:lambda}
\end{figure*}

The DEMIST method consists of a hyperparameter $\lambda$ that penalizes the loss of task-specific features while performing denoising (Eq.~\eqref{eq:hybrid_loss}). To qualitatively demonstrate the effect of this parameter, we present a representative result in Fig.~\ref{fig:lambda}. To generate this result, we denoised an MPI-SPECT image in the test set acquired at low-dose level of 6.25\% with trained DEMIST networks associated with varying $\lambda$ values. We observe that, for this example, assigning a higher weight to the task-specific loss term (as achieved by increasing $\lambda$) leads to improved defect visibility in the denoised image. This improvement indicates that the incorporation of task-specific loss term preserves features used by human observers for performing detection tasks. These results also illustrate that the $\lambda$ parameter can be interpreted as a term that controls the smoothness in the image. A lower $\lambda$ value results in an increased weight for the fidelity term, and is observed to lead to increased blur in the image which then translates to the defect being washed out.

We note in Figs.~\ref{fig:auc}-\ref{fig:auc_def_sev_cho} that while DEMIST yielded improved performance on detection task compared to low-dose and TADL approaches, there is room for improvement compared to normal-dose protocol. To further improve performance, a more advanced network architecture \cite{fan2022sunet} with the proposed task-specific loss term could be used. Also, given the heterogeneity in patient characteristics, increasing the amount of training data may make the method generalize well to test data and thus improve performance \cite{chaudhary2022fast}. However, there is a possibility of fundamental information loss that might not be retrievable even if we increase the amount of training data. This topic requires further investigation. Moreover, we considered three low-dose levels but there may be other low-dose levels for which the proposed method may yield performance that is similar to normal-dose protocol.

In this paper, we developed the task-specific denoising method in the context of cardiac SPECT. However, the method is general and could be applied to other medical imaging modalities where the task of interest is detecting abnormalities. Other applications could include reducing administered radiation dose in oncological PET images and reducing acquisition time for oncological magnetic resonance (MR) images. Also, our method was developed in the context of detection tasks, but could be applicable for other tasks where the task performance depends on mathematical features that could be extracted from the images.    

Our study has several limitations. The first limitation is that DEMIST was validated with model observes and not human observers. While we considered a CHO-based model observer that has been shown to emulate human-observer performance, conducting this study with human observers would provide a more rigorous validation of the method. Additionally, the signal location was known to the \rnew{anthropomorphic }observer, but in clinical settings, this location is not known. \rchange{A human-observer study}{Furthermore, given the location-known settings, we could not assess the performance of the method on falsely detecting defects at other locations. Localization ROC studies with human observers} will enable \rnew{us to }\rchange{validating}{validate} whether the proposed method \rchange{could}{can} improve human-observer performance on the task of perfusion-defect detection with unknown defect location. Reliable performance in a human-observer study would provide confidence for the clinical translation of this method. 
Here we point out that to test the robustness of the method to different
channelized observers, we also conducted the evaluation with another observer, namely the channelized multi-template observer \cite{li2016use}. Our findings, which are provided in the Supplementary material and in \cite{rahman2023demist-supp-github}, show that even with this observer, DEMIST significantly outperformed low-dose protocol and TADL method.
This finding shows the robustness of the method to different observers. 

A second limitation of this study is that the DEMIST method was trained with data where the defect-present cases contained synthetic inserted defects. This was because, during training, the knowledge of the presence of defect and the location of defect centroid in defect-present cases was required. Ideally, though, DEMIST should be trained using data with real perfusion defects. However, determining the ground truth regarding the presence of defects and their centroid is challenging. To address a similar issue of lack of ground truth while training a network to delineate tumors in PET images, Leung \etal pre-trained a network with multiple simulated images where the tumor boundaries were known exactly, and then fine-tuned with a small number of clinical images \cite{leung2020physics}. A similar strategy of pre-training DEMIST with multiple synthetic-defect images and then fine-tuning this network with a small number of training images where the defect centroid is obtained manually \rchange{could be investigated}{presents an area of future study}. Another limitation was that we considered defects in only two regions. Increasing the number of defect locations, including septal and lateral walls of the LV, would provide further insights on the robustness of the proposed method. \rchange{Finally}{Furthermore, in practical scenarios, there could be multiple defect locations in the same case. The proposed DEMIST method can be extended by extracting channel vectors from each of these locations. Additionally}, the method was evaluated with only single-center data. However, the results motivate evaluation of the data across multiple centers to assess the generalizability of this technique across centers.\rnew{ Finally, the method was developed for non-gated MPI SPECT images. Another area of future research is advancing this method to gated MPI SPECT \cite{paul2004gated,zhang2023assessment}. One challenge here is identifying the center of the defect. The proposed method can be advanced to account for this issue by extracting channel vectors for a neighborhood of possible defect centers.} 

%====================================================================================
% Conclusions
%====================================================================================
\section{CONCLUSION}
A task-specific deep-learning-based method (DEMIST) was proposed to denoise low-dose MPI-SPECT images with the goal of improving performance on the clinical task of detecting perfusion defects compared to low-dose images. For this purpose, we introduced a task-specific loss term in our loss function that penalizes the loss of anthropomorphic channel features.
%, which are expected to assist in  the detection task based on our understanding of the human visual system and prior literature on model-observer studies. 
%Evaluation with retrospective clinical MPI-SPECT studies demonstrated that the DEMIST method improved performance on the task of detecting perfusion defects with an anthropomorphic channelized Hotelling observer. 
According to the RELAINCE guidelines \cite{jha2022nuclear}, our evaluation study yields the following claim: \textit{A deep-learning-based task-specific denoising method for MPI-SPECT improved performance of images acquired at 6.25\%, 12.5\% and 25\% dose levels on the task of detecting inserted location-known perfusion defects with a significance level of 5\% as evaluated in a retrospective clinical study with single-center multi-scanner data and with an anthropomorphic channelized Hotelling observer.} The results provide strong evidence to evaluate DEMIST with human observers. 
%====================================================================================
% Ack
%====================================================================================
\section{ACKNOWLEDGMENT}
This work was supported by the National Institute of Biomedical Imaging and Bioengineering of the National Institute of Health under grants R01-EB031051, R01-EB031962 and the NSF CAREER award. We also thank the Research Infrastructure Service (RIS) at Washington University for providing computational resources.
%====================================================================================
% Reference
%====================================================================================
\IEEEtriggercmd{\enlargethispage{-5in}}
\bibliographystyle{IEEEtran}
\bibliography{ref_opt}

%====================================================================================
\clearpage
\setcounter{section}{0}
\setcounter{figure}{0}
\setcounter{table}{0}
\renewcommand{\thesection}{S-\Roman{section}}
\renewcommand{\thefigure}{S-\arabic{figure}}
\renewcommand{\thetable}{S-\arabic{table}}
\onecolumn

\begin{center}
    \Huge Supplementary Materials
\end{center}

\section{Network architecture}\label{sec:supp_network}
\begin{table}[!ht]
    \centering
    \caption{Network architecture (Conv. = convolutional, BN = Batch Normalization, ReLU = Rectified Linear Unit, MaxPool = Max Pooling).}
    \begin{tabular}{|c|c|c|c|c|c|c|}
    \hline
        \textbf{Layer} & \textbf{Layer type} & \textbf{Number of filters} & \textbf{Filter size} & \textbf{Stride/Pool size} & \textbf{Input size} & \textbf{Output size} \\ \hline
        1 & Conv. & 16 & 3×3×3 & 1×1×1 & 48×48×48×1 & 48×48×48×16 \\ \hline
        2 & BN + Leaky ReLU & - & - & - & 48×48×48×16 & 48×48×48×16 \\ \hline
        3 & MaxPool & - & - & 2×2×2 & 24×24×24×16 & 24×24×24×16 \\ \hline
        4 & Conv. & 32 & 3×3×3 & 1×1×1 & 24×24×24×16 & 24×24×24×32 \\ \hline
        5 & BN + Leaky ReLU & - & - & - & 24×24×24×32 & 24×24×24×32 \\ \hline
        6 & MaxPool & - & - & 2×2×2 & 12×12×12×32 & 12×12×12×32 \\ \hline
        7 & Conv. & 64 & 3×3×3 & 1×1×1 & 12×12×12×32 & 12×12×12×64 \\ \hline
        8 & BN + Leaky ReLU & - & - & - & 12×12×12×64 & 12×12×12×64 \\ \hline
        9 & MaxPool & - & - & 2×2×2 & 12×12×12×64 & 6×6×6×64 \\ \hline
        10 & Conv. & 128 & 3×3×3 & 1×1×1 & 6×6×6×64 & 6×6×6×128 \\ \hline
        11 & BN + Leaky ReLU & - & - & - & 6×6×6×128 & 6×6×6×128 \\ \hline
        12 & Dropout & - & - & - & 6×6×6×128 & 6×6×6×128 \\ \hline
        13 & Transposed Conv. & 64 & 3×3×3 & 2×2×2 & 6×6×6×128 & 12×12×12×64 \\ \hline
        14 & BN + Leaky ReLU & - & - & - & 12×12×12×64 & 12×12×12×64 \\ \hline
        15 & Add Layer 8 & - & - & - & 12×12×12×64 & 12×12×12×64 \\ \hline
        16 & Conv. & 64 & 3×3×3 & 1×1×1 & 12×12×12×64 & 12×12×12×64 \\ \hline
        17 & BN + Leaky ReLU & - & - & - & 12×12×12×64 & 12×12×12×64 \\ \hline
        18 & Transposed Conv. & 32 & 3×3×3 & 2×2×2 & 12×12×12×64 & 24×24×24×32 \\ \hline
        19 & BN + Leaky ReLU & - & - & - & 24×24×24×32 & 24×24×24×32 \\ \hline
        20 & Add Layer 5 & - & - & - & 24×24×24×32 & 24×24×24×32 \\ \hline
        21 & Conv. & 32 & 3×3×3 & 1×1×1 & 24×24×24×32 & 24×24×24×32 \\ \hline
        22 & BN + Leaky ReLU & - & - & - & 24×24×24×32 & 24×24×24×32 \\ \hline
        23 & Transposed Conv. & 16 & 3×3×3 & 2×2×2 & 24×24×24×32 & 48×48×48×16 \\ \hline
        24 & BN + Leaky ReLU & - & - & - & 48×48×48×16 & 48×48×48×16 \\ \hline
        25 & Add Layer 2 & - & - & - & 48×48×48×16 & 48×48×48×16 \\ \hline
        26 & Conv. & 16 & 3×3×3 & 1×1×1 & 48×48×48×16 & 48×48×48×16 \\ \hline
        27 & BN + Leaky ReLU & - & - & - & 48×48×48×16 & 48×48×48×16 \\ \hline
        28 & Conv & 1 & 1×1×1 & 1×1×1 & 48×48×48×1 & 48×48×48×1 \\ \hline
        29 & BN + Leaky ReLU & - & - & - & 48×48×48×1 & 48×48×48×1 \\ \hline
    \end{tabular}
\end{table}

\clearpage
\section{Evaluation with channelized multi-template observer}\label{sec:supp_mto}
\subsection{Background}
To assess the robustness of DEMIST method across various channelized model observers, we performed our objective evaluation study using a different channelized model observer, namely a channelized multi-template observer (CMTO). In our population, the  defect sizes, severities and locations were all varying. In this case, it has been observed that the channel outputs (vectors) for the entire population may not be multivariate normally distributed \cite{elshahaby2016factors}, thus limiting the applicability of the widely known Hotelling observer. However, the channel outputs for sub-ensembles of patient data grouped based on defect type may have multivariate normal distributions \cite{li2016use}. For this case, a CMTO was developed and evaluated in the context of MPI SPECT \cite{li2016use}. The CMTO applies the Hotelling template to the channel outputs and adds a constant term to compute test statistics for each sub-ensemble, and calculates a single global area under the ROC curve (AUC) using the pooled test statistics from all the sub-ensembles. The observer yields the maximal AUC when shifting the distributions of Hotelling observer test statistics by a different constant for each sub-ensemble is allowed \cite{li2016use}. The channels chosen for this observer were also the rotationally symmetric frequency channels, as in the CHO study. Also, the clinical task for this observer was detecting perfusion defects where the defect location was known.

\subsection{Generation of test statistic}
We followed the same procedure as described in Sec.~III-C of the manuscript to obtain channel vectors. For each sub-ensemble, the channel vectors of defect-present and defect-absent population were used to learn the template using a leave-one-out approach. The collection of test statistic from each sub-ensemble were then pooled \cite{li2016use} and were used to perform the ROC analysis.

%\clearpage
\subsection{Results}
\begin{figure}[H]
\centering
\includegraphics[width = 0.45\linewidth]{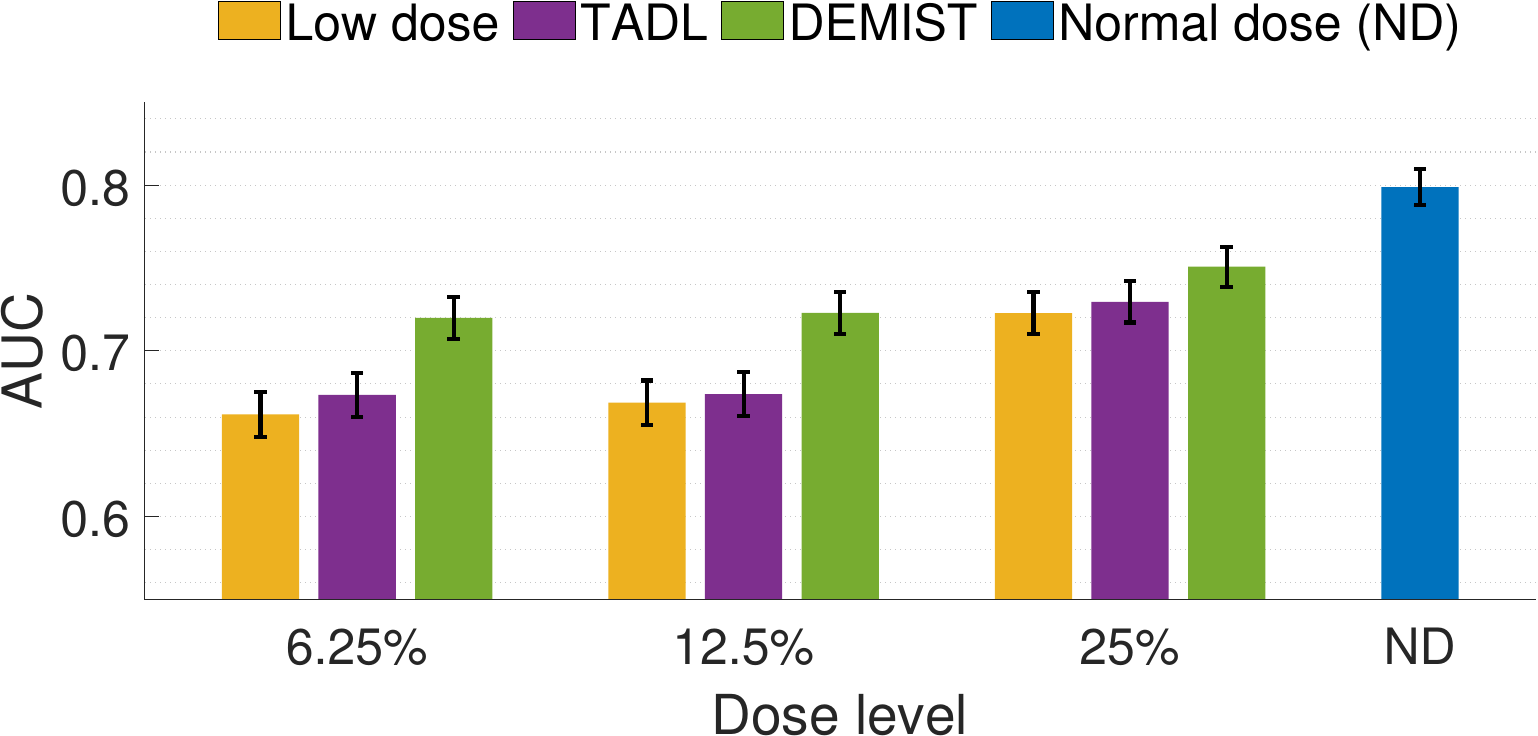}
\caption{AUC values obtained for the normal and low-dose images, the images denoised using the proposed DEMIST approach and TADL approach at various dose levels with CMTO. Error bars denote 95\% confidence intervals.
}
\label{fig:auc_mto}
\end{figure}

Fig.~\ref{fig:auc_mto} shows the AUC values obtained with the low-dose protocol, TADL method, DEMIST method and normal-dose protocol. We observed that at all dose levels, the DEMIST method  yielded a significant improvement ($p<0.05$) in performance on detection task compared to the low-dose protocol as well as the TADL method. The TADL method generally did not improve performance compared to the low-dose protocol.
%\subsubsection{Non-stratified analysis}

\begin{figure}[H]
\centering
\includegraphics[width = 0.45\linewidth]{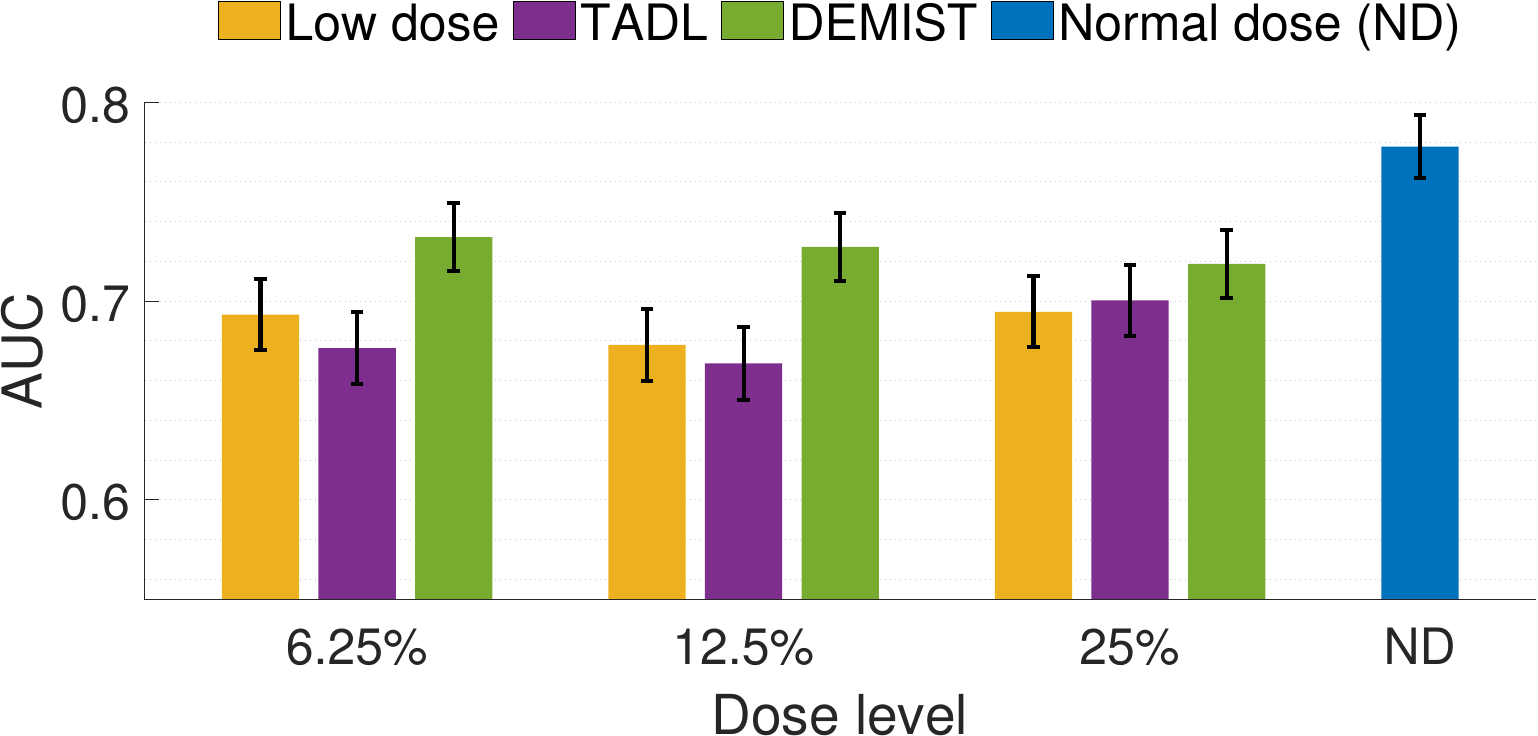}
\caption{AUC values for male patients obtained for the different approaches and at various dose levels using CMTO. Error bars denote 95\% confidence intervals.
}
\label{fig:auc_male_mto}
\end{figure}

\begin{figure}[H]
\centering
\includegraphics[width = 0.45\linewidth]{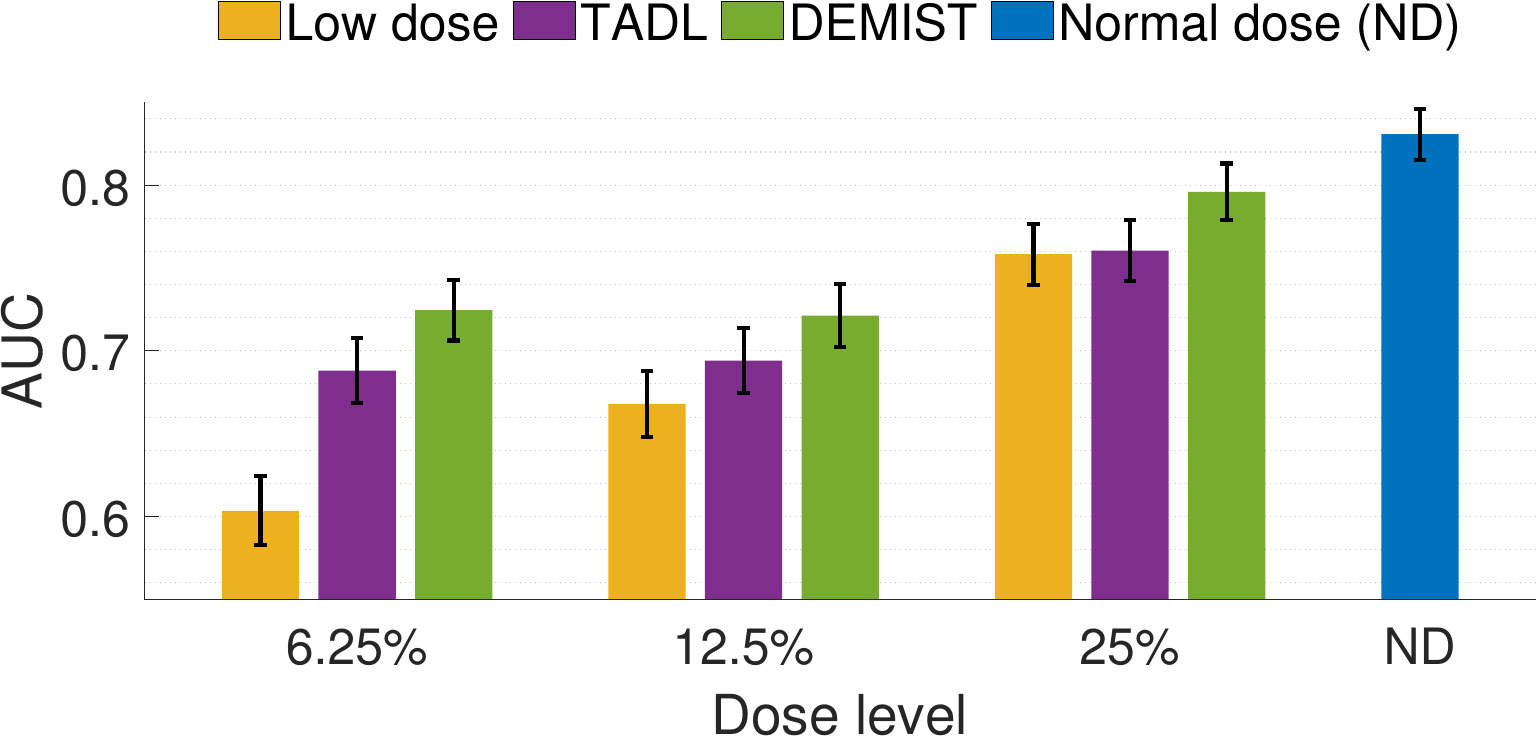}
\caption{AUC values obtained for the different approaches and at various dose levels with female patients using CMTO. Error bars denote 95\% confidence intervals.
}
\label{fig:auc_female_mto}
\end{figure}

Figs.~\ref{fig:auc_male_mto} and \ref{fig:auc_female_mto} show the AUC values obtained for stratified analysis based on sex. We observed that at all dose levels and stratified groups, the DEMIST method  yielded a significant improvement ($p<0.05$) in performance on detection task compared to the low-dose protocol and TADL method. Again, we observed that the TADL method generally did not improve performance significantly compared to the low-dose protocol.  

\begin{figure*}[h]
\centering
\includegraphics[width = 0.9\linewidth]{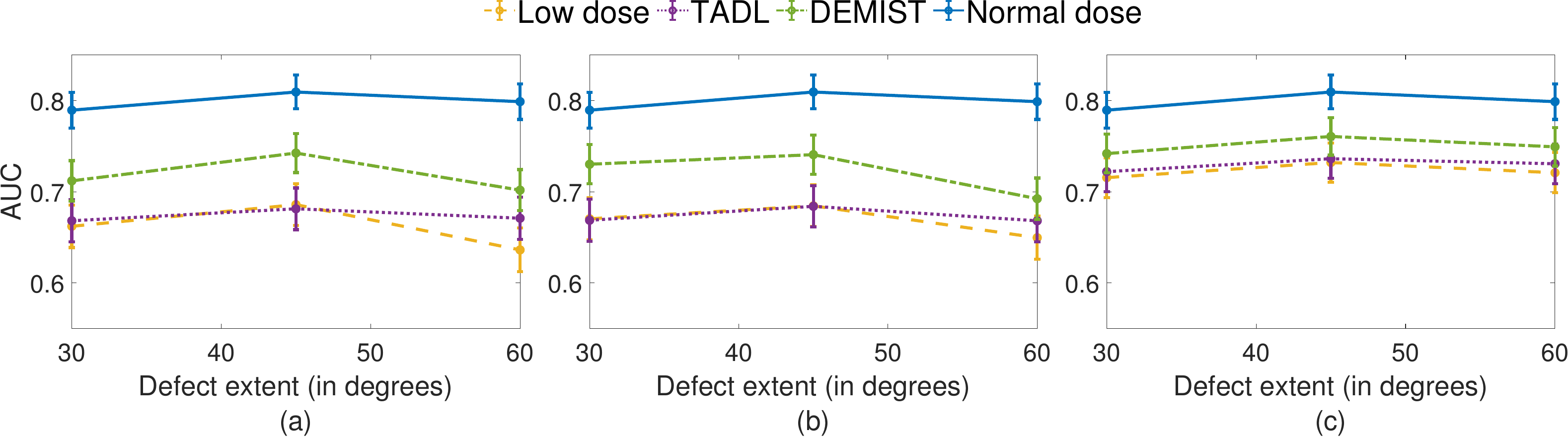}
\caption{AUC values obtained using CMTO for the various approaches as a function of different defect extents with (a) 6.25\%, (b) 12.5\% and (c) 25\% dose levels. Error bars denote 95\% confidence intervals.
}
\label{fig:auc_def_ext_mto}
\end{figure*}

\begin{figure*}[h]
\centering
\includegraphics[width = 0.9\linewidth]{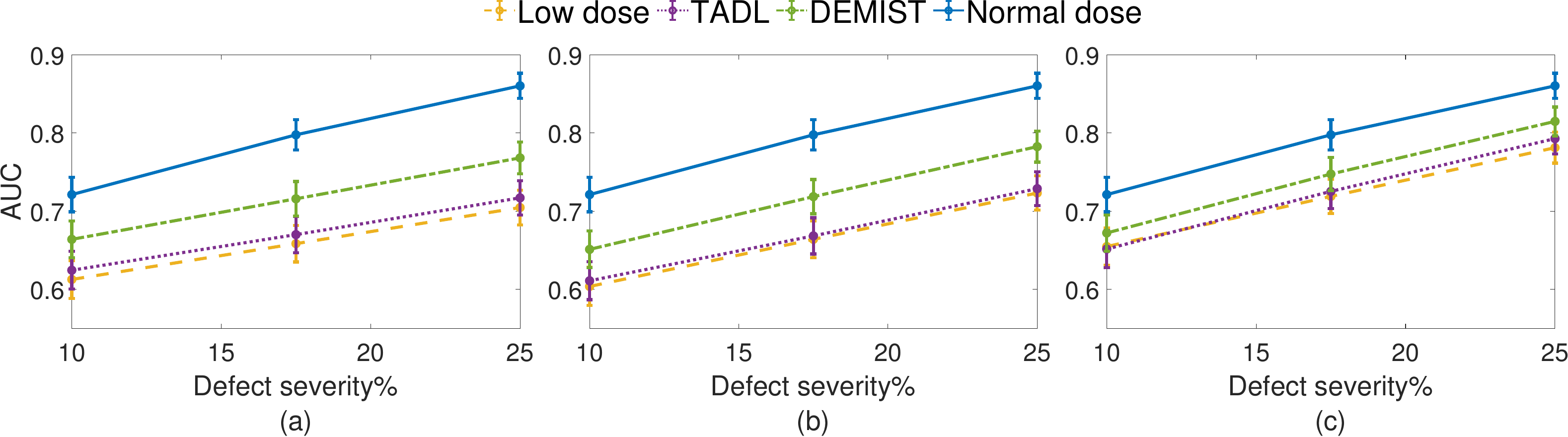}
\caption{AUC values obtained using CMTO for the various approaches as a function of different defect severities with (a) 6.25\%, (b) 12.5\% and (c) 25\% dose levels. Error bars denote 95\% confidence intervals.
}
\label{fig:auc_def_sev_mto}
\end{figure*}

Figs.~\ref{fig:auc_def_ext_mto} and \ref{fig:auc_def_sev_mto} show the AUC values obtained for stratified analysis based on defect extent and severity, respectively. Similar to previous results, we observed that at all dose levels and stratified groups, the DEMIST method  yielded a significant improvement ($p<0.05$) in performance on detection task compared to low-dose protocol and TADL method. Again, the TADL method generally did not improve performance significantly compared to the low-dose protocol.

\begin{figure*}[h]
\centering
\includegraphics[width = 0.9\linewidth]{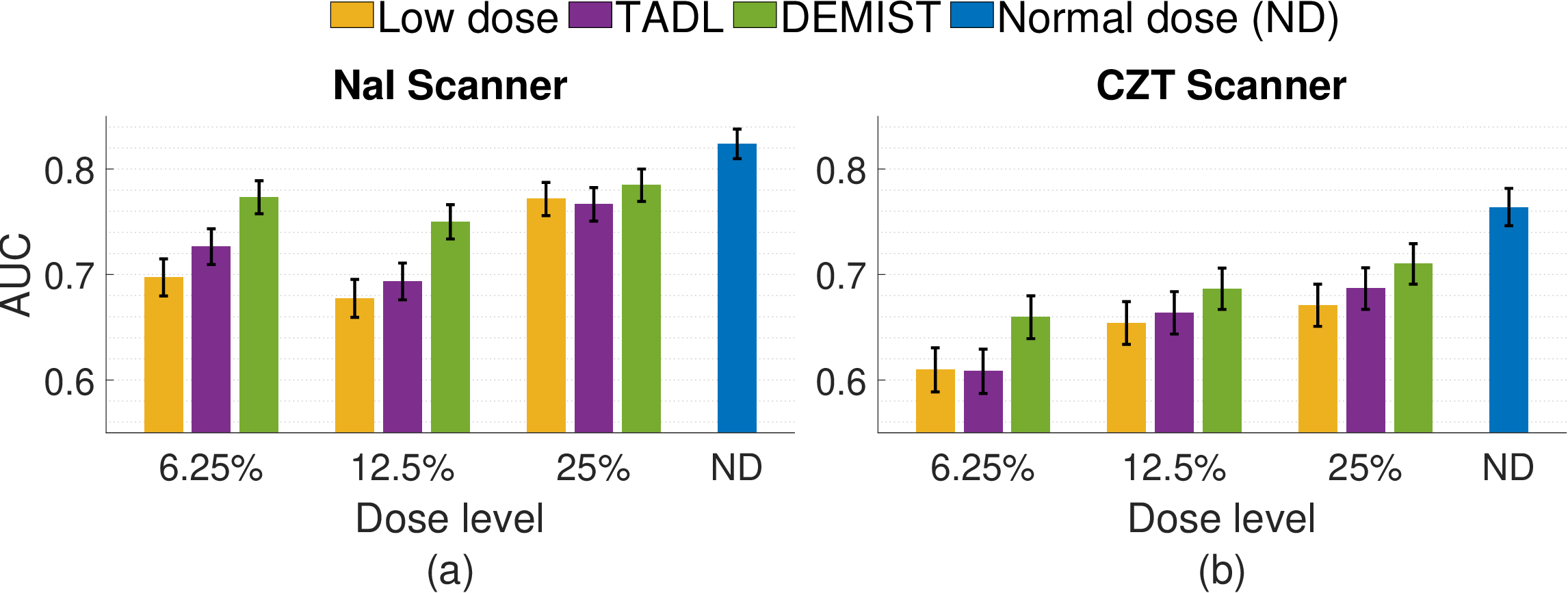}
\caption{AUC values obtained using CMTO for the various approaches for (a) NaI and (b) CZT scanner at various dose levels. Error bars denote 95\% confidence intervals.
}
\label{fig:auc_scanner_mto}
\end{figure*}

Fig.~\ref{fig:auc_scanner_mto} shows the AUC values obtained for stratified analysis based on scanner type. We observed that at all dose levels, for both NaI and CZT scanners, the DEMIST method  yielded a significant improvement ($p<0.05$) in performance on detection task compared to low-dose protocol and TADL method. The TADL method generally did not improve performance significantly compared to the low-dose protocol.

%\FloatBarrier 

%\subsubsection{Stratified analysis based on sex}
%\FloatBarrier 

%\subsubsection{Stratified analysis based on defect extent}

%\subsubsection{Stratified analysis based on defect severity}

\clearpage
\section{Corrected $p$-values of Delong’s test for AUC comparison between two methods}\label{sec:supp_pval}
% \usepackage{multirow}
\begin{table}[h!]
\centering
\footnotesize
\caption{Corrected p-values for various analyses and observers (LD = low-dose protocol)}
\label{tab:my-table}
\begin{tabular}{|c|c|c|c|lll|}
\hline
\multirow{2}{*}{Analysis type} &
  \multirow{2}{*}{Stratified group} &
  \multirow{2}{*}{Observer} &
  \multirow{2}{*}{Comparisons} &
  \multicolumn{3}{c|}{Dose levels} \\ \cline{5-7} 
 &                            &                      &                & \multicolumn{1}{c|}{6.25\%}   & \multicolumn{1}{c|}{12.50\%}  & \multicolumn{1}{c|}{25\%} \\ \hline
\multirow{6}{*}{Non-stratified analysis} &
  \multirow{6}{*}{-} &
  \multirow{3}{*}{CMTO} &
  DEMIST VS LD &
  \multicolumn{1}{l|}{2.72E-32} &
  \multicolumn{1}{l|}{8.05E-38} &
  1.44E-28 \\ \cline{4-7} 
 &                            &                      & DEMIST VS TADL & \multicolumn{1}{l|}{2.74E-29} & \multicolumn{1}{l|}{1.19E-29} & 2.08E-14                  \\ \cline{4-7} 
 &                            &                      & TADL VS LD     & \multicolumn{1}{l|}{0.111834} & \multicolumn{1}{l|}{0.840879} & 0.079587                  \\ \cline{3-7} 
 &                            & \multirow{3}{*}{CHO} & DEMIST VS LD   & \multicolumn{1}{l|}{8.78E-20} & \multicolumn{1}{l|}{5.72E-10} & 2.64E-17                  \\ \cline{4-7} 
 &                            &                      & DEMIST VS TADL & \multicolumn{1}{l|}{5.87E-11} & \multicolumn{1}{l|}{3.71E-17} & 6.22E-08                  \\ \cline{4-7} 
 &                            &                      & TADL VS LD     & \multicolumn{1}{l|}{0.014445} & \multicolumn{1}{l|}{1.634715} & 0.035604                  \\ \hline
 &                            &                      &                & \multicolumn{1}{l|}{}         & \multicolumn{1}{l|}{}         &                           \\ \hline
\multirow{13}{*}{Stratified analysis based on sex} &
  \multirow{6}{*}{Male} &
  \multirow{3}{*}{CMTO} &
  DEMIST VS LD &
  \multicolumn{1}{l|}{4.43E-07} &
  \multicolumn{1}{l|}{1.83E-15} &
  4.8E-09 \\ \cline{4-7} 
 &                            &                      & DEMIST VS TADL & \multicolumn{1}{l|}{2.33E-18} & \multicolumn{1}{l|}{4.39E-18} & 0.000553                  \\ \cline{4-7} 
 &                            &                      & TADL VS LD     & \multicolumn{1}{l|}{0.130437} & \multicolumn{1}{l|}{0.626301} & 0.607878                  \\ \cline{3-7} 
 &                            & \multirow{3}{*}{CHO} & DEMIST VS LD   & \multicolumn{1}{l|}{0.002403} & \multicolumn{1}{l|}{4.71E-05} & 7.44E-07                  \\ \cline{4-7} 
 &                            &                      & DEMIST VS TADL & \multicolumn{1}{l|}{8.25E-10} & \multicolumn{1}{l|}{0.000263} & 0.018342                  \\ \cline{4-7} 
 &                            &                      & TADL VS LD     & \multicolumn{1}{l|}{0.130599} & \multicolumn{1}{l|}{1.255887} & 0.350955                  \\ \cline{2-7} 
 &                            &                      &                & \multicolumn{1}{l|}{}         & \multicolumn{1}{l|}{}         &                           \\ \cline{2-7} 
 & \multirow{6}{*}{Female}    & \multirow{3}{*}{CMTO} & DEMIST VS LD   & \multicolumn{1}{l|}{2.59E-49} & \multicolumn{1}{l|}{1.21E-16} & 1.13E-26                  \\ \cline{4-7} 
 &                            &                      & DEMIST VS TADL & \multicolumn{1}{l|}{5.46E-05} & \multicolumn{1}{l|}{1.04E-05} & 3.56E-17                  \\ \cline{4-7} 
 &                            &                      & TADL VS LD     & \multicolumn{1}{l|}{6.51E-15} & \multicolumn{1}{l|}{0.001287} & 1.792665                  \\ \cline{3-7} 
 &                            & \multirow{3}{*}{CHO} & DEMIST VS LD   & \multicolumn{1}{l|}{1.51E-18} & \multicolumn{1}{l|}{3.47E-07} & 2.46E-13                  \\ \cline{4-7} 
 &                            &                      & DEMIST VS TADL & \multicolumn{1}{l|}{0.132264} & \multicolumn{1}{l|}{1.27E-17} & 1.94E-08                  \\ \cline{4-7} 
 &                            &                      & TADL VS LD     & \multicolumn{1}{l|}{8.02E-11} & \multicolumn{1}{l|}{0.200286} & 0.41967                   \\ \hline
 &                            &                      &                & \multicolumn{1}{l|}{}         & \multicolumn{1}{l|}{}         &                           \\ \hline
\multirow{20}{*}{Stratified analysis based on defect extent} &
  \multirow{6}{*}{30$^{\circ}$ extent} &
  \multirow{3}{*}{CMTO} &
  DEMIST VS LD &
  \multicolumn{1}{l|}{2.08E-08} &
  \multicolumn{1}{l|}{4.02E-15} &
  1.04E-09 \\ \cline{4-7} 
 &                            &                      & DEMIST VS TADL & \multicolumn{1}{l|}{7.07E-10} & \multicolumn{1}{l|}{5.29E-14} & 4.73E-05                  \\ \cline{4-7} 
 &                            &                      & TADL VS LD     & \multicolumn{1}{l|}{1.581237} & \multicolumn{1}{l|}{2.56014}  & 0.606618                  \\ \cline{3-7} 
 &                            & \multirow{3}{*}{CHO} & DEMIST VS LD   & \multicolumn{1}{l|}{1.4E-05}  & \multicolumn{1}{l|}{0.025992} & 1.57E-06                  \\ \cline{4-7} 
 &                            &                      & DEMIST VS TADL & \multicolumn{1}{l|}{2.42E-05} & \multicolumn{1}{l|}{0.000472} & 0.00423                   \\ \cline{4-7} 
 &                            &                      & TADL VS LD     & \multicolumn{1}{l|}{1.077156} & \multicolumn{1}{l|}{2.296377} & 0.362862                  \\ \cline{2-7} 
 &                            &                      &                & \multicolumn{1}{l|}{}         & \multicolumn{1}{l|}{}         &                           \\ \cline{2-7} 
 & \multirow{6}{*}{45$^{\circ}$ extent} & \multirow{3}{*}{CMTO} & DEMIST VS LD   & \multicolumn{1}{l|}{1.1E-11}  & \multicolumn{1}{l|}{1.08E-15} & 9.54E-13                  \\ \cline{4-7} 
 &                            &                      & DEMIST VS TADL & \multicolumn{1}{l|}{1.37E-17} & \multicolumn{1}{l|}{4.72E-15} & 1.53E-06                  \\ \cline{4-7} 
 &                            &                      & TADL VS LD     & \multicolumn{1}{l|}{1.904022} & \multicolumn{1}{l|}{2.836206} & 1.246464                  \\ \cline{3-7} 
 &                            & \multirow{3}{*}{CHO} & DEMIST VS LD   & \multicolumn{1}{l|}{2.75E-09} & \multicolumn{1}{l|}{2.93E-05} & 0.000266                  \\ \cline{4-7} 
 &                            &                      & DEMIST VS TADL & \multicolumn{1}{l|}{5.04E-07} & \multicolumn{1}{l|}{4.47E-08} & 0.004428                  \\ \cline{4-7} 
 &                            &                      & TADL VS LD     & \multicolumn{1}{l|}{0.681075} & \multicolumn{1}{l|}{1.96272}  & 2.055168                  \\ \cline{2-7} 
 &                            &                      &                & \multicolumn{1}{l|}{}         & \multicolumn{1}{l|}{}         &                           \\ \cline{2-7} 
 & \multirow{6}{*}{60$^{\circ}$ extent} & \multirow{3}{*}{CMTO} & DEMIST VS LD   & \multicolumn{1}{l|}{1.31E-13} & \multicolumn{1}{l|}{4.45E-08} & 4.55E-08                  \\ \cline{4-7} 
 &                            &                      & DEMIST VS TADL & \multicolumn{1}{l|}{7.56E-05} & \multicolumn{1}{l|}{0.002286} & 0.000327                  \\ \cline{4-7} 
 &                            &                      & TADL VS LD     & \multicolumn{1}{l|}{0.002466} & \multicolumn{1}{l|}{0.130788} & 0.251469                  \\ \cline{3-7} 
 &                            & \multirow{3}{*}{CHO} & DEMIST VS LD   & \multicolumn{1}{l|}{1.08E-10} & \multicolumn{1}{l|}{2.12E-06} & 2.58E-08                  \\ \cline{4-7} 
 &                            &                      & DEMIST VS TADL & \multicolumn{1}{l|}{0.088839} & \multicolumn{1}{l|}{3.15E-07} & 0.001602                  \\ \cline{4-7} 
 &                            &                      & TADL VS LD     & \multicolumn{1}{l|}{0.000753} & \multicolumn{1}{l|}{1.500372} & 0.080946                  \\ \hline
 &                            &                      &                & \multicolumn{1}{l|}{}         & \multicolumn{1}{l|}{}         &                           \\ \hline
\multirow{20}{*}{Stratified analysis based on defect severity} &
  \multirow{6}{*}{10\% severity} &
  \multirow{3}{*}{CMTO} &
  DEMIST VS LD &
  \multicolumn{1}{l|}{1.11E-07} &
  \multicolumn{1}{l|}{1.21E-07} &
  0.001269 \\ \cline{4-7} 
 &                            &                      & DEMIST VS TADL & \multicolumn{1}{l|}{4.07E-07} & \multicolumn{1}{l|}{6.94E-06} & 0.000468                  \\ \cline{4-7} 
 &                            &                      & TADL VS LD     & \multicolumn{1}{l|}{0.784584} & \multicolumn{1}{l|}{1.289538} & 1.785861                  \\ \cline{3-7} 
 &                            & \multirow{3}{*}{CHO} & DEMIST VS LD   & \multicolumn{1}{l|}{0.000543} & \multicolumn{1}{l|}{0.038628} & 0.001143                  \\ \cline{4-7} 
 &                            &                      & DEMIST VS TADL & \multicolumn{1}{l|}{0.142011} & \multicolumn{1}{l|}{0.021213} & 0.055386                  \\ \cline{4-7} 
 &                            &                      & TADL VS LD     & \multicolumn{1}{l|}{0.346419} & \multicolumn{1}{l|}{2.395917} & 0.808731                  \\ \cline{2-7} 
 &                            &                      &                & \multicolumn{1}{l|}{}         & \multicolumn{1}{l|}{}         &                           \\ \cline{2-7} 
 &
  \multirow{6}{*}{17.5\% severity} &
  \multirow{3}{*}{CMTO} &
  DEMIST VS LD &
  \multicolumn{1}{l|}{2.88E-11} &
  \multicolumn{1}{l|}{1.56E-13} &
  2.73E-10 \\ \cline{4-7} 
 &                            &                      & DEMIST VS TADL & \multicolumn{1}{l|}{2.83E-10} & \multicolumn{1}{l|}{5.92E-11} & 9.09E-06                  \\ \cline{4-7} 
 &                            &                      & TADL VS LD     & \multicolumn{1}{l|}{0.719325} & \multicolumn{1}{l|}{1.793304} & 0.692793                  \\ \cline{3-7} 
 &                            & \multirow{3}{*}{CHO} & DEMIST VS LD   & \multicolumn{1}{l|}{2.35E-07} & \multicolumn{1}{l|}{0.000505} & 9.99E-07                  \\ \cline{4-7} 
 &                            &                      & DEMIST VS TADL & \multicolumn{1}{l|}{0.00026}  & \multicolumn{1}{l|}{1.07E-06} & 0.001629                  \\ \cline{4-7} 
 &                            &                      & TADL VS LD     & \multicolumn{1}{l|}{0.312813} & \multicolumn{1}{l|}{2.061603} & 0.514611                  \\ \cline{2-7} 
 &                            &                      &                & \multicolumn{1}{l|}{}         & \multicolumn{1}{l|}{}         &                           \\ \cline{2-7} 
 &
  \multirow{6}{*}{25\% severity} &
  \multirow{3}{*}{CMTO} &
  DEMIST VS LD &
  \multicolumn{1}{l|}{7.94E-16} &
  \multicolumn{1}{l|}{5.31E-21} &
  1.33E-17 \\ \cline{4-7} 
 &                            &                      & DEMIST VS TADL & \multicolumn{1}{l|}{3.04E-14} & \multicolumn{1}{l|}{1.6E-15}  & 3.68E-07                  \\ \cline{4-7} 
 &                            &                      & TADL VS LD     & \multicolumn{1}{l|}{0.542664} & \multicolumn{1}{l|}{1.39266}  & 0.038385                  \\ \cline{3-7} 
 &                            & \multirow{3}{*}{CHO} & DEMIST VS LD   & \multicolumn{1}{l|}{1.58E-12} & \multicolumn{1}{l|}{4.38E-07} & 6.98E-10                  \\ \cline{4-7} 
 &                            &                      & DEMIST VS TADL & \multicolumn{1}{l|}{1.13E-09} & \multicolumn{1}{l|}{3.32E-13} & 6.31E-05                  \\ \cline{4-7} 
 &                            &                      & TADL VS LD     & \multicolumn{1}{l|}{0.260334} & \multicolumn{1}{l|}{1.118241} & 0.340011                  \\ \hline
\end{tabular}
\end{table}
\begin{table}[]
\caption{Corrected p-values for stratified analysis with scanner types}
\label{tab:my-table-scan}
\begin{tabular}{|c|c|c|c|ccc|}
\hline
\multirow{2}{*}{Analysis type} &
  \multirow{2}{*}{Stratified group} &
  \multirow{2}{*}{Observer} &
  \multirow{2}{*}{Comparisons} &
  \multicolumn{3}{c|}{Dose levels} \\ \cline{5-7} 
 &  &                      &                & \multicolumn{1}{c|}{6.25\%}   & \multicolumn{1}{c|}{12.50\%}  & 25\%     \\ \hline
\multirow{13}{*}{Stratified analysis based on scanner} &
  \multirow{6}{*}{NaI} &
  \multirow{3}{*}{CMTO} &
  DEMIST VS LD &
  \multicolumn{1}{c|}{1.79E-38} &
  \multicolumn{1}{c|}{4.33E-37} &
  3.59E-06 \\ \cline{4-7} 
 &  &                      & DEMIST VS TADL & \multicolumn{1}{c|}{1.50E-21} & \multicolumn{1}{c|}{2.24E-21} & 4.51E-07 \\ \cline{4-7} 
 &  &                      & TADL VS LD     & \multicolumn{1}{c|}{8.97E-05} & \multicolumn{1}{c|}{0.05029}  & 0.428633 \\ \cline{3-7} 
 &  & \multirow{3}{*}{CHO} & DEMIST VS LD   & \multicolumn{1}{c|}{1.79E-14} & \multicolumn{1}{c|}{1.59E-08} & 2.554406 \\ \cline{4-7} 
 &  &                      & DEMIST VS TADL & \multicolumn{1}{c|}{0.00164}  & \multicolumn{1}{c|}{0.454727} & 0.057514 \\ \cline{4-7} 
 &  &                      & TADL VS LD     & \multicolumn{1}{c|}{4.93E-06} & \multicolumn{1}{c|}{0.000214} & 0.066835 \\ \cline{2-7} 
 &  &                      &                & \multicolumn{1}{c|}{}         & \multicolumn{1}{c|}{}         &          \\ \cline{2-7} 
 &
  \multirow{6}{*}{CZT} &
  \multirow{3}{*}{CMTO} &
  DEMIST VS LD &
  \multicolumn{1}{c|}{6.16E-08} &
  \multicolumn{1}{c|}{4.68E-06} &
  4.50E-13 \\ \cline{4-7} 
 &  &                      & DEMIST VS TADL & \multicolumn{1}{c|}{1.48E-08} & \multicolumn{1}{c|}{0.00161}  & 3.14E-05 \\ \cline{4-7} 
 &  &                      & TADL VS LD     & \multicolumn{1}{c|}{2.692152} & \multicolumn{1}{c|}{0.64725}  & 0.018067 \\ \cline{3-7} 
 &  & \multirow{3}{*}{CHO} & DEMIST VS LD   & \multicolumn{1}{c|}{0.00263}  & \multicolumn{1}{c|}{1.830386} & 1.88E-16 \\ \cline{4-7} 
 &  &                      & DEMIST VS TADL & \multicolumn{1}{c|}{0.000259} & \multicolumn{1}{c|}{5.87E-11} & 4.20E-05 \\ \cline{4-7} 
 &  &                      & TADL VS LD     & \multicolumn{1}{c|}{2.378452} & \multicolumn{1}{c|}{6.44E-06} & 0.000155 \\ \hline
\end{tabular}
\end{table}
\clearpage
\section{SPECT scanner configuration}\label{sec:supp_scanner}
% \usepackage{multirow}

\iffalse
\begin{table}[h!]
\centering
\caption{Acquisition and reconstruction parameters of SPECT/CT systems. (LEHR=Low-energy high-resolution)}
\label{tab:my-table}
\begin{tabular}{|c|cc|}
\hline
\multirow{2}{*}{Properties}      & \multicolumn{2}{c|}{Scanners}                             \\ \cline{2-3} 
                                 & \multicolumn{1}{c|}{GE 670 CZT}    & GE Discovery 670 Pro \\ \hline
Collimator type                  & \multicolumn{1}{c|}{LEHR}          & LEHR                 \\ \hline
Collimator grid                  & \multicolumn{1}{c|}{Parallel hole} & Parallel hole        \\ \hline
Photopeak energy window (in keV) & \multicolumn{1}{c|}{126-154}       & 126-154              \\ \hline
Reconstruction                   & \multicolumn{1}{c|}{OSEM}          & OSEM                 \\ \hline
Subsets                          & \multicolumn{1}{c|}{6}             & 6                    \\ \hline
Iteration                        & \multicolumn{1}{c|}{8}             & 8                    \\ \hline
Attenuation correction           & \multicolumn{1}{c|}{CT}            & CT                   \\ \hline
\end{tabular}
\end{table}
\fi

% \usepackage{multirow}
\begin{table}[h!]
\centering
\caption{Acquisition and reconstruction parameters of SPECT/CT systems. (LEHR=Low-energy high-resolution,WEHR=Wide-energy high-resolution)}
\label{tab:my-table}
\begin{tabular}{|c|cc|}
\hline
\multirow{2}{*}{}                     & \multicolumn{2}{c|}{Scanner}                       \\ \cline{2-3} 
                                 & \multicolumn{1}{c|}{GE Discovery NM/CT 670 Pro NaI} & GE Discovery NM/CT 670 Pro CZT \\ \hline
Number of cases in test data          & \multicolumn{1}{c|}{63}            & 51            \\ \hline
Number of cases in train data         & \multicolumn{1}{c|}{102}           & 82            \\ \hline
Number of cases in validation data    & \multicolumn{1}{c|}{12}            & 28            \\ \hline
Collimator type                       & \multicolumn{1}{c|}{LEHR}          & WEHR          \\ \hline
Collimator grid                       & \multicolumn{1}{c|}{Parallel hole} & Parallel hole \\ \hline
Detector                              & \multicolumn{1}{c|}{NaI}           & CZT           \\ \hline
Energy resolution at 140 keV (\%)     & \multicolumn{1}{c|}{9.8}           & 6.3           \\ \hline
Intrinsic spatial resolution (in mm)  & \multicolumn{1}{c|}{3.9}           & 2.46          \\ \hline
System sensitivity (cps/MBq) at 10 cm & \multicolumn{1}{c|}{72}         & 85            \\ \hline
Photopeak energy window (in keV) & \multicolumn{1}{c|}{126-154}                        & 126-154                        \\ \hline
Reconstruction                        & \multicolumn{1}{c|}{OSEM}          & OSEM          \\ \hline
Subsets                               & \multicolumn{1}{c|}{6}             & 6             \\ \hline
Iteration                             & \multicolumn{1}{c|}{8}             & 8             \\ \hline
Attenuation correction                & \multicolumn{1}{c|}{CT}            & CT            \\ \hline
\end{tabular}
\end{table}

\clearpage
\section{Examples of defect signals}\label{sec:def_signals}
\begin{figure*}[h]
\centering
\includegraphics[width = 0.9\linewidth]{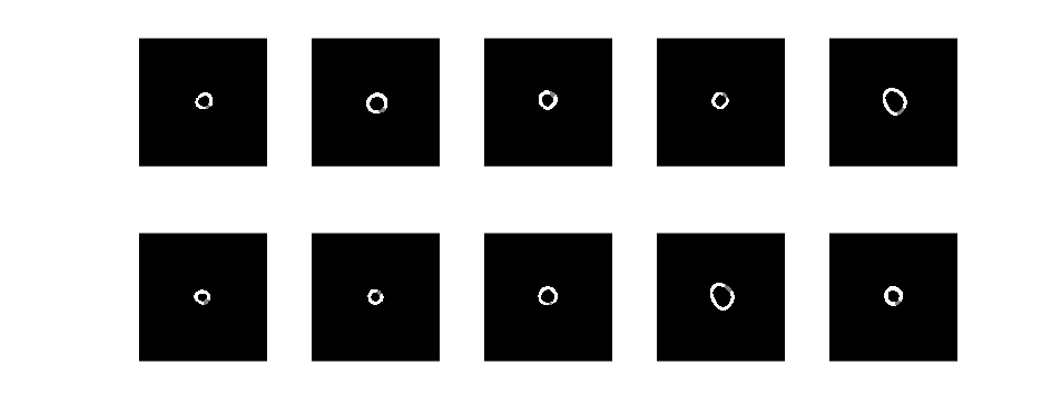}
\caption{Examples of inserted defect using the LV segmented mask with varying extents and locations. The defects are at 50\% severity for illustration purpose. 
}
\label{fig:def_signals}
\end{figure*}

\end{document}